\documentclass[aps,showpacs,preprintnumbers,amsmath,amssymb]{revtex4}
\usepackage{dcolumn}
\usepackage[dvips]{epsfig}
\usepackage{graphicx}
\usepackage{amssymb}
\usepackage{color}
\usepackage{enumerate}
\usepackage{subfigure}

\begin{document}
\baselineskip=0.8 cm
%\preprint{example}
\title{Shadow of a Schwarzschild black hole surrounded by a Bach-Weyl ring}
%\altaffiliation{}
\author{Mingzhi Wang$^{1,2}$,  Songbai Chen$^{1,3}$\footnote{Corresponding author: csb3752@hunnu.edu.cn}, Jieci Wang$^{1}$ \footnote{jcwang@hunnu.edu.cn}, Jiliang
Jing$^{1,3}$ \footnote{jljing@hunnu.edu.cn}
}
\affiliation{ $ ^1$ Department of Physics, Key Laboratory of Low Dimensional Quantum Structures
and Quantum Control of Ministry of Education, Synergetic Innovation Center for Quantum Effects and Applications, Hunan
Normal University,  Changsha, Hunan 410081, People's Republic of China
\\
$ ^2$School of Mathematics and Physics, Qingdao University of Science and Technology, Qingdao, Shandong 266061, People's Republic of China
 \\
$ ^3$Center for Gravitation and Cosmology, College of Physical Science and Technology,
Yangzhou University, Yangzhou 225009, People's Republic of China}

\begin{abstract}
\baselineskip=0.6 cm
\begin{center}
{\bf Abstract}
\end{center}

We have studied the shadows of a Schwarzschild black hole surrounded by a Bach-Weyl ring through the backward ray-tracing method. The presence of Bach-Weyl ring leads to that the photon dynamical system is non-integrable and then chaos would appear in the photon motion, which affects sharply the  black hole shadow. The size and shape the black hole shadow depend on the black hole parameter, the Bach-Weyl ring mass and the Weyl radius between black hole and ring. Some self-similar fractal structures also appear in the black hole shadow, which originates from the chaotic lensing. We also study the change of the image of Bach-Weyl ring with the ring mass and the Weyl radius. Finally, we analyze the invariant manifolds of Lyapunov orbits near the fixed points and discuss further the formation of the shadow of a Schwarzschild black hole with Bach-Weyl ring.

\end{abstract}
\maketitle

\newpage
\section{Introduction}

The first image of the supermassive black hole in the center of the giant elliptical galaxy M87 has been obtained by using Event Horizon Telescope \cite{fbhs1,fbhs2,fbhs3,fbhs4,fbhs5,fbhs6}, which is one of the most exciting events in astrophysics and black hole physics since it provides the first direct visual evidence  that there exists exactly black hole in our Universe. The information stored in this image can help us to understand further shadow of black hole, the matter accretion process and black hole jets.
A black hole shadow is a two-dimensional dark region in the observer's sky, which is caused by light rays that fall into an event horizon when propagated backwards in time. Since the shape and size of shadow carry the fingerprint of the geometry around the black hole \cite{sha2,sha3}, the shadow can be regarded as a potential tool to probe black hole parameters. For example, the shadow of a Schwarzschild black hole is a perfect black disk,  but it becomes an elongated silhouette for a Kerr-like black hole due to the dragging effect arising from the black hole rotation \cite{sha2,sha3}. Moreover, the recent investigations have shown that the cusp silhouette of shadow appears for a Kerr black hole with Proca hair \cite{fpos2} and for a Konoplya-Zhidenko rotating non-Kerr black hole \cite{sb10} if the spacetime parameters lie in a certain range. Interestingly, there exist self-similar
fractal structures in the black hole shadow for the cases where the photon motion is non-integrable,
such as, a rotating black hole with scalar hair \cite{sw,swo,astro,chaotic}, a Majumdar-Papapetrou binary black hole system \cite{binary, sha18},  Bonnor black diholes with magnetic dipole moment \cite{my}, and a non-Kerr rotating compact object with quadrupole mass moment \cite{sMN}. These novel structure and patterns also appear in the shadows of a two dynamic black hole system with the positive cosmological constant \cite{colliding}. It is found that these self-similar
fractal structures and patterns in black hole shadows are determined actually by the non-planar bound photon orbits \cite{fpos2} and  the invariant phase space structures \cite{BI} of the photon motion in the black hole spacetimes. The recent investigations on the shadows of black holes characterizing by other parameters \cite{sha4,sha5,sha6,sha7,sha8,sha9,sha10,sha11,sha12,sha13,sha14,sha141,sha15,sha16,
sb1,sha17,sha19,sha191,sha192,sha193,sha194,shan1,shan1add,shan2add,shan3add} indicate that these black hole parameters yield the richer structure and patterns for shadows in various theories of gravity.

In general, a real black hole in the galaxy owns an accretion disk. This extra gravitational source around the black hole also affect the propagation of photon and further modify the shape of black hole shadow. Considering the radiation from a thin accretion disk around a black hole, Luminet obtained a simulated photograph of a Schwarzschild black hole with thin accretion disk \cite{grr0}, which shows that  there are primary and secondary images of the thin accretion disk outside black hole shadow. The image for a Kerr black hole with Keplerian accretion disk has been simulated in Refs. \cite{grr,short,BKD}.  With the general relativistic ray-tracing code, F. H. Vincent et al \cite{gyoto} has studied the images of a thin infinite accretion disk and Ion torus around compact objects. These investigations are very important to understand the effects of accretion disk on the black hole images. It is well known that the photon motion is non-integrable in the gravity system of a black hole with extra gravitational sources including accretion disk. However, it is still an open issue how the non-integrable of photon motion arising from extra gravitational sources affect the black hole shadows. The main purpose of this paper is to answer it through studying the shadow of a Schwarzschild black hole surrounded by Bach-Weyl ring.
The Bach-Weyl ring is a concentric thin ring described by the Bach-Weyl solution \cite{byr}, which can be looked as a general relativistic analog of a Newtonian ring of constant density. For a Schwarzschild black hole with Bach-Weyl ring, the presence of the ring changes the spacetime structure and affects geodesic motion of particle in the spacetime \cite{ong, structure, Geodesics}. It is shown that there exist  chaotic motion for particle with certain initial conditions in this background spacetime \cite{free}. In this paper, we will study the shadow of Schwarzschild black hole with Bach-Weyl ring and then probe the effects of the photon chaotic motion on the shadow.

The paper is organized as follows. In Sec. II, we review briefly the spacetime of a Schwarzschild black hole surrounded by Bach-Weyl ring and then analyze the null geodesics equation in this spacetime. In Sec. III, we adopt the backward ray-tracing technique and present numerically the shadows for the Schwarzschild black hole surrounded by Bach-Weyl ring. In Sec.IV, we analyze the invariant phase-space structures and further explain the formation of shadow cast by a Schwarzschild black hole with a Bach-Weyl ring. Finally, we present a summary.

\section{The spacetime of a Schwarzschild black hole with a Bach-Weyl ring and null geodesics}

The spacetime of a vacuum static and axially symmetric spacetime,  in general, can be described by the Weyl metric
\begin{eqnarray}
\label{zjzdg}
ds^{2}=-e^{2\nu}dt^{2}+e^{2\lambda-2\nu}
(d\rho^{2}+dz^{2})+\rho^{2}e^{-2\nu}d\phi^{2},
\end{eqnarray}
where $\nu$ and $\lambda$ only are the functions of $\rho$ and $z$.  The function $\nu(\rho, z)$ satisfies the Laplace equation and can be superposed linearly, which behaves like the gravitational potential in the Newtonian theory. However, the quantity $\lambda$ does not own such a property of  linear superposition. For a gravity system containing two gravitational sources with their individual functions $\nu_{1}$, $\lambda_{1}$ and $\nu_{2}$, $\lambda_{2}$, one can write the functions $\nu$ and  $\lambda$ for the whole system as $\nu=\nu_{1}+\nu_{2}$ and $\lambda=\lambda_{1}+\lambda_{2}+\lambda_{int}$, respectively. The quantity  $\lambda_{int}$ is the interaction term which obeys the equations \cite{ong}
\begin{eqnarray}
\label{xhzyx}
&&\lambda_{int,\rho}=2\rho(\nu_{1,\rho}\nu_{2,\rho}-\nu_{1,z}\nu_{2,z})\\ \nonumber
&&\lambda_{int,z}=2\rho(\nu_{1,\rho}\nu_{2,z}+\nu_{1,z}\nu_{2,\rho}).
\end{eqnarray}
Here, we focus on the spacetime generated
by a Schwarzschild-type black hole with a
thin ring described by the Bach-Weyl solution. In the Schwarzschild coordinates, the gravity of such a spacetime can be described by the metric \cite{byr, ong, free}
\begin{eqnarray}
\label{bydg}
ds^{2}=-(1-\frac{2M}{r})e^{2\nu_{BW}}dt^{2}+\frac{e^{2\lambda_{ext}-2\nu_{BW}}}{1-\frac{2M}{r}}dr^{2}+r^{2}e^{-2\nu_{BW}}(e^{2\lambda_{ext}}d\theta^{2}+\sin^{2}\theta d\phi^{2}),
\end{eqnarray}
where $\lambda_{ext}=\lambda_{BW}+\lambda_{int}$. The functions $\nu_{BW}$ and $\lambda_{BW}$ have the forms \cite{byr, ong}
\begin{eqnarray}
\label{bydgxs}
&&\nu_{BW}=-\frac{2\mathcal{M}K(k)}{\pi l_{2}},\\ \nonumber
&&\lambda_{BW}=-\frac{\mathcal{M}^{2}}{4\pi^{2}b^{2}\rho}\bigg[(\rho+b)(E-K)^{2}+\frac{(\rho-b)(E-k^{'2}K)^{2}}{k^{'2}}\bigg],
\end{eqnarray}
where $\mathcal{M}$ is the mass of Bach-Weyl ring, $b$ is the Weyl radius and
$l_{1,2}=\sqrt{(\rho\mp b)^{2}+z^{2}}$.
$K(k)$ and $E(k)$ are the 1st and the 2nd kind complete elliptic integrals with the forms
\begin{eqnarray}
\label{KE}
&&K(k)=\int_{0}^{\pi/2}\frac{d\alpha}{\sqrt{1-k^{2}\sin^{2}\alpha}},\\ \nonumber
&&E(k)=\int_{0}^{\pi/2}\sqrt{1-k^{2}\sin^{2}\alpha}d\alpha,
\end{eqnarray}
and
\begin{eqnarray}
\label{KExs}
k^{2}=1-\frac{l_{1}^{2}}{l_{2}^{2}}=\frac{4\rho b}{l_{2}^{2}},\;\;\;\;\;\;\;\;\;\;k^{'2}=1-k^{2}=\frac{l_{1}^{2}}{l_{2}^{2}}.
\end{eqnarray}
For a Schwarzschild black hole with a Bach-Weyl ring, the function $\lambda_{int}$  has not an analytical form, which means that the form of $\lambda_{int}$ can be obtained only by resorting to numerical computations.

With the metric (\ref{bydg}), one can find that the Hamiltonian of a photon propagation in the spacetime of a Schwarzschild black hole surrounded by a Bach-Weyl ring can be expressed as
\begin{equation}
\label{bwhami}
H(x,p)=-\frac{re^{-2\nu_{BW}}}{2(r-2M)}p_{t}^{2}
+\frac{e^{2\nu_{BW}}(r-2M)}{2re^{2\lambda_{ext}}}
p_{r}^{2}+\frac{e^{2\nu_{BW}}}{2r^{2}e^{2\lambda_{ext}}}
p_{\theta}^{2}+\frac{e^{2\nu_{BW}}}{2r^{2}\sin^{2}\theta}p_{\phi}^{2}.
\end{equation}
It is obvious that this Hamiltonian is not an explicit function of time coordinate $t$ and angular coordinate $\phi$. Therefore, there exist two 	integration constants $E$ and $L_{z}$ for the null geodesics motion
\begin{eqnarray}
\label{EL}
E=-p_{t}=-g_{tt}\dot{t}=\frac{(r-2M)e^{2\nu_{BW}}}{r}\dot{t},\;\;\;\;\;\;\;\;\;\;\;\;
L_{z}=p_{\phi}=g_{\phi\phi}\dot{\phi}=\frac{r^{2}\sin^{2}\theta}{e^{2\nu_{BW}}}\dot{\phi}.
\end{eqnarray}
Making use of these two conserved quantities, we find the equations of photon motion in this spacetime with a Bach-Weyl ring can be simplified as
\begin{eqnarray}
\label{bwcdx}
\dot{t}&&=\frac{Ere^{-2\nu_{BW}}}{r-2M},\\ \nonumber
\ddot{r}&&=\frac{e^{-4\lambda_{ext}}}{r^{4}(r-2M)}\bigg\{2e^{-4\lambda_{ext}}(r-2M)r^{4}\dot{r}\dot{\theta}\nu_{BW,\theta}-e^{2\lambda_{ext}}E^{2}r^{3}\bigg[M\\ \nonumber&&-(r^{2}-2Mr)\nu_{BW,r}\bigg]+e^{4\lambda_{ext}}r^{3}\bigg[M\dot{r}^{2}+r(r-2M)^{2}\dot{\theta}^{2}+(r^{2}\\ \nonumber&&-2Mr)\bigg(2\dot{r}\dot{\theta}\lambda_{ext,\theta}+(r^{2}\dot{\theta}^{2}-2Mr\dot{\theta}^{2}-\dot{r}^{2})(\nu_{BW,r}-\lambda_{ext,r})\bigg)\bigg]\\ \nonumber&&+e^{4\nu_{BW}}L_{z}(r-2M)^{2}\csc^{2}\theta(r\lambda_{ext,r}-r\nu_{BW,r}+1)\bigg\}, \\ \nonumber
\ddot{\theta}&&=\frac{e^{-4\lambda_{ext}}}{r^{4}(r-2M)}\bigg\{-e^{4\lambda_{ext}}E^{2}r^{3}\nu_{BW,\theta}+e^{4\nu_{BW}}L_{z}^{2}(r-2M)\csc^{2}\theta(\cot\theta\\ \nonumber&&+\lambda_{ext,\theta}-\nu_{BW,\theta})
+e^{4\lambda_{ext}}r^{3}\bigg[(r^{2}\dot{\theta}^{2}-2Mr\dot{\theta}^{2}-\dot{r}^{2})(\lambda_{ext,\theta}-\nu_{BW,\theta})\\ \nonumber&&-2(r-2M)\dot{r}\dot{\theta}(r\lambda_{ext,r}-r\nu_{BW,r}+1)\bigg]\bigg\},\\ \nonumber
\dot{\phi}&&=\frac{L_{z}e^{2\nu_{BW}}}{r^{2}\sin^{2}\theta}.
\end{eqnarray}
Due to the Bach-Weyl ring, one can find the $r-$ component and $\theta-$ component do not decouple from each other, which means that the dynamical system of photon in this spacetime is non-integrable and then chaos could occur in this case. It implies that the presence of Bach-Weyl ring will bring
some novel features of the black hole shadow.

\section{Shadow casted by a Schwarzschild black hole surrounded by a Bach-Weyl ring}

In this section, we adopt the backward ray-tracing technique \cite{sw,swo,astro,chaotic,binary,sha18,my,sMN,BI,swo7} to simulate numerically the shadow of a Schwarzschild black hole surrounded by a Bach-Weyl ring.  Here, it is assumed that the light rays evolve from the observer backward in time and then the information carried by each ray would be respectively assigned to a pixel in a final image in the observer's sky. Along this spirit, we solve numerically the null geodesic equation (\ref{bwcdx}) for each pixel in the final image with the corresponding initial condition and obtain the image of shadow in observer's sky which is composed of the pixels connected to the light rays falling down into the horizon of black hole.
Since the spacetime (\ref{bydg}) is asymptotically flat,  one can expand the observer basis $\{e_{\hat{t}},e_{\hat{r}},e_{\hat{\theta}},e_{\hat{\phi}}\}$ as a form in the coordinate basis
\cite{sw,swo,astro,chaotic,binary,sha18,my,sMN,BI,swo7}
\begin{eqnarray}
\label{zbbh}
e_{\hat{\mu}}=e^{\nu}_{\hat{\mu}} \partial_{\nu},
\end{eqnarray}
where the transform matrix $e^{\nu}_{\hat{\mu}}$ satisfies $g_{\mu\nu}e^{\mu}_{\hat{\alpha}}e^{\nu}_{\hat{\beta}}
=\eta_{\hat{\alpha}\hat{\beta}}$, and $\eta_{\hat{\alpha}\hat{\beta}}$ is the usual Minkowski metric.
Considering that the spacetime of a Schwarzschild black hole with a Bach-Weyl ring (\ref{bydg}) is static and axially symmetric, it is convenient to choice a decomposition
\begin{eqnarray}
\label{zbbh1}
e^{\nu}_{\hat{\mu}}=\left(\begin{array}{cccc}
\frac{1}{\sqrt{-g_{00}}}&0&0&0\\
0&\frac{1}{\sqrt{g_{11}}}&0&0\\
0&0&\frac{1}{\sqrt{g_{22}}}&0\\
0&0&0&\frac{1}{\sqrt{g_{33}}}
\end{array}\right).
\end{eqnarray}
And then, the locally measured four-momentum $p^{\hat{\mu}}$ of a photon can be obtained through the projection of its four-momentum $p^{\mu}$  onto $e_{\hat{\mu}}$, i.e.,
\begin{eqnarray}
\label{dl}
p^{\hat{t}}&=&-p_{\hat{t}}=-e^{\nu}_{\hat{t}} p_{\nu}=\sqrt{\frac{r}{r-2M}}e^{-\nu_{BW}}E, \nonumber\\
p^{\hat{r}}&=&p_{\hat{r}}=e^{\nu}_{\hat{r}} p_{\nu}=\sqrt{\frac{r-2M}{r}}e^{\nu_{BW}-\lambda_{ext}}p_{r} ,\nonumber\\
p^{\hat{\theta}}&=&p_{\hat{\theta}}=e^{\nu}_{\hat{\theta}} p_{\nu}=\frac{e^{\nu_{BW}-\lambda_{ext}}}{r}p_{\theta},
\nonumber\\
p^{\hat{\phi}}&=&p_{\hat{\phi}}=e^{\nu}_{\hat{\phi}} p_{\nu}=\frac{e^{\nu_{BW}}}{r\sin\theta}L_z.
\end{eqnarray}
\begin{figure}
\subfigure[$\mathcal{M}=0.5M, b=10M$]{ \includegraphics[width=5.4cm ]{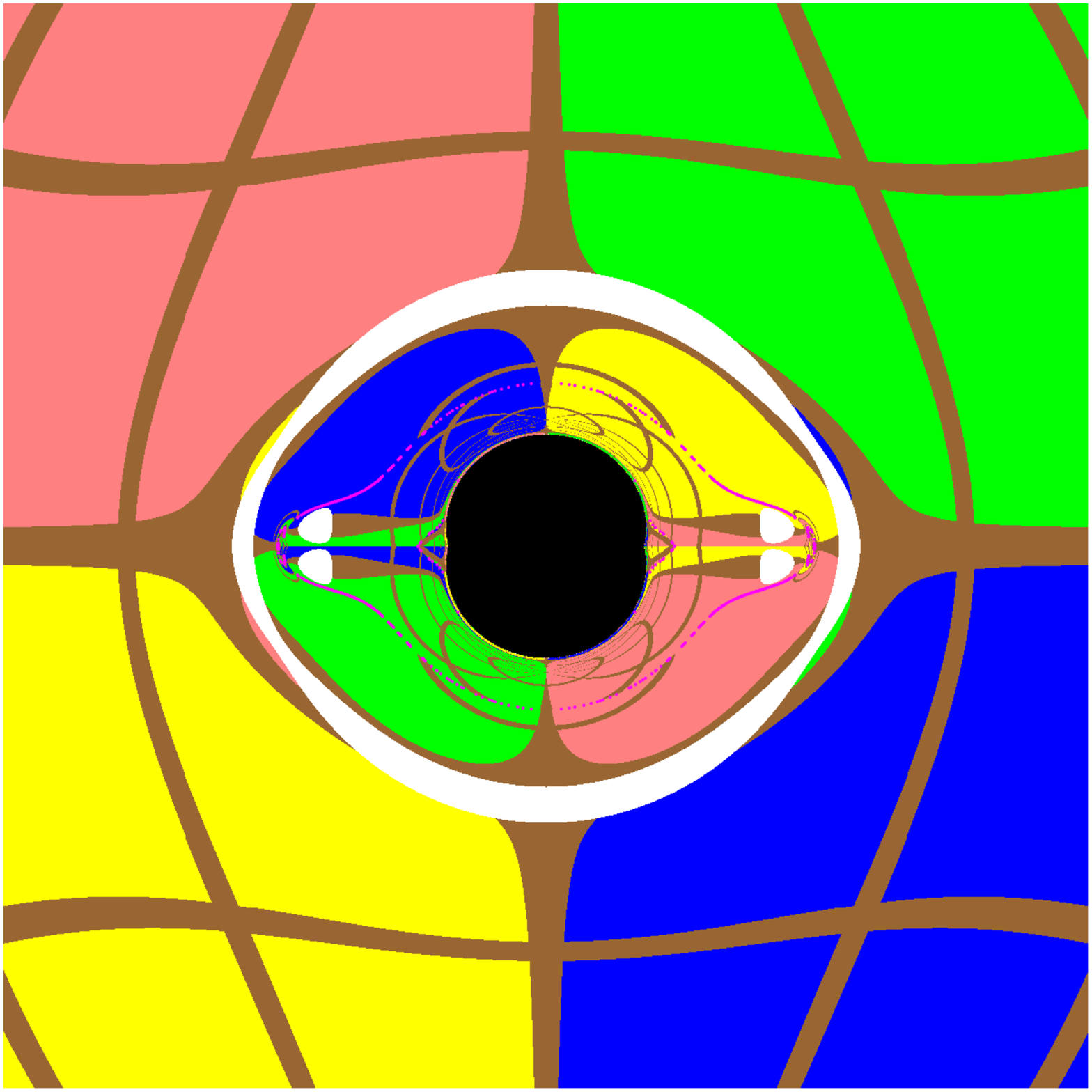}}\subfigure[$\mathcal{M}=1M, b=10M$]{ \includegraphics[width=5.4cm ]{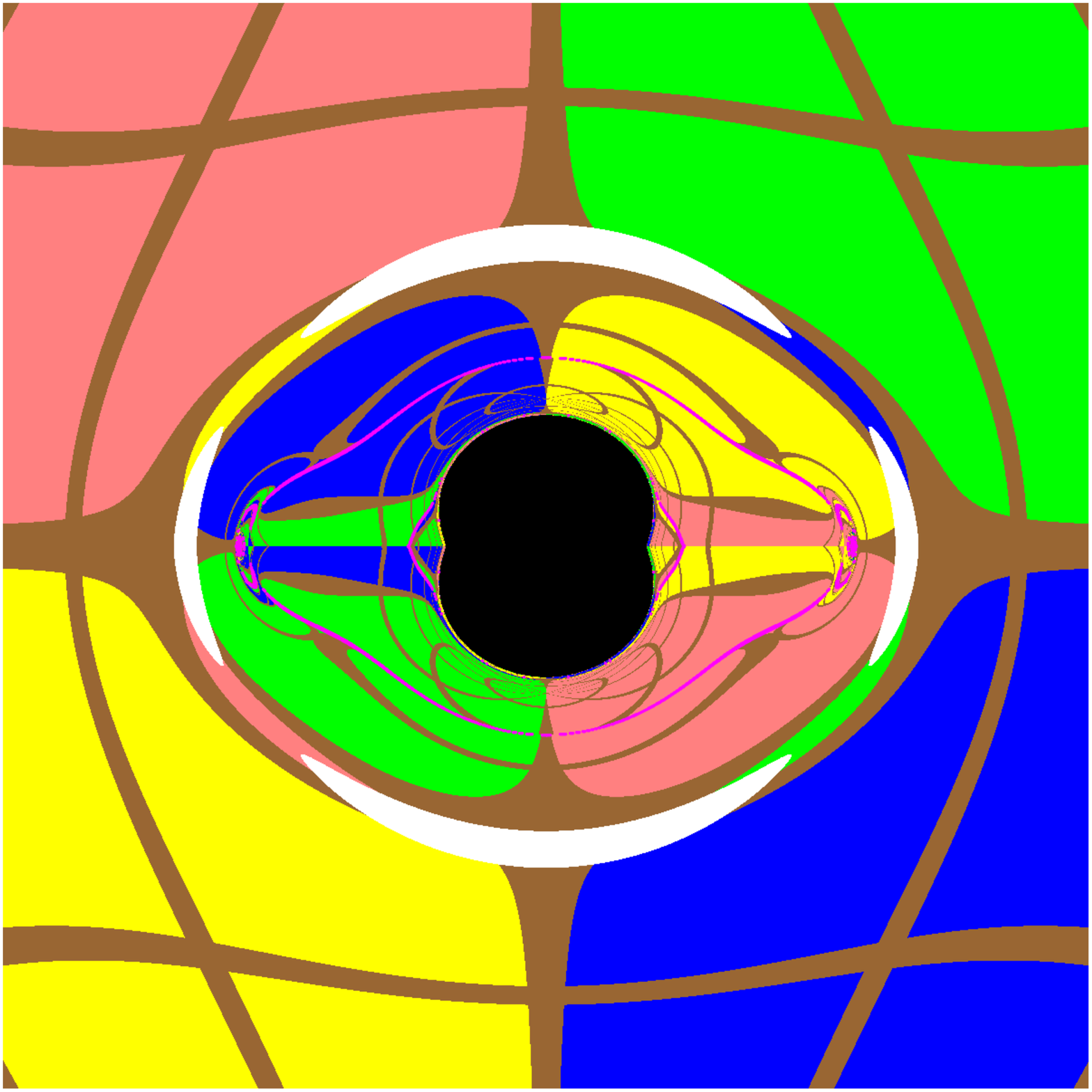}} \subfigure[$\mathcal{M}=1.5M, b=10M$]{ \includegraphics[width=5.4cm ]{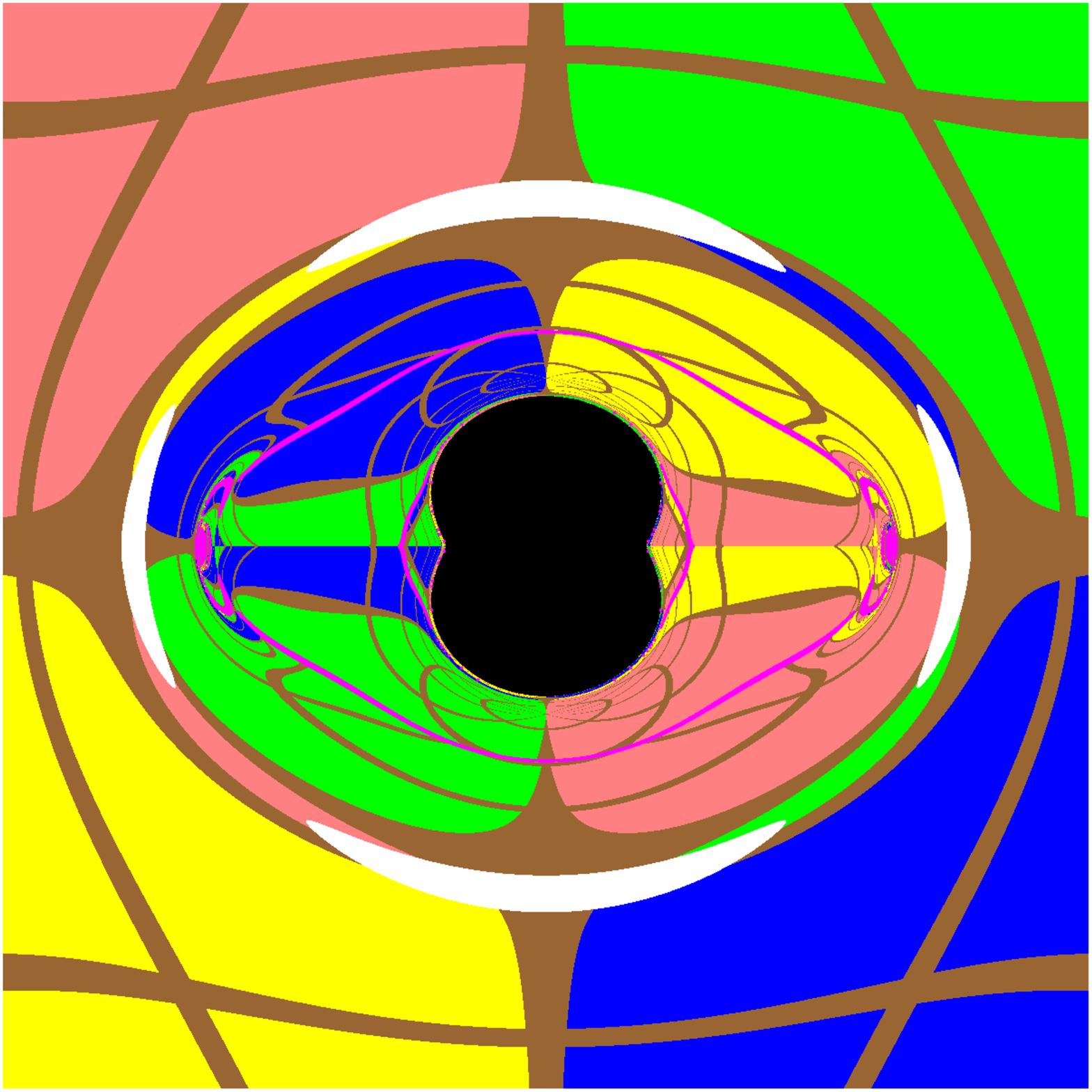}}
\subfigure[$\mathcal{M}=0.5M, b=20M$]{ \includegraphics[width=5.4cm ]{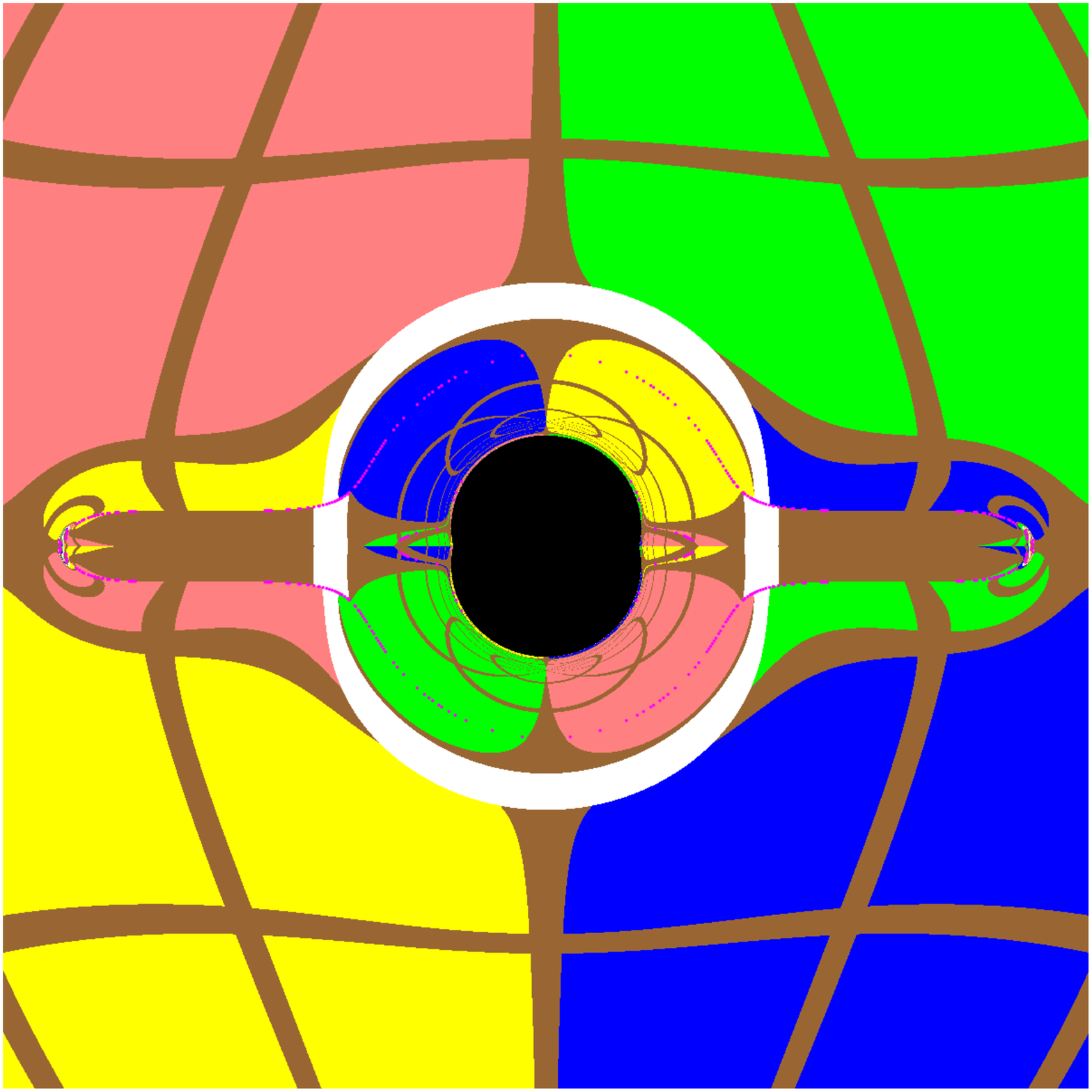}}\subfigure[$\mathcal{M}=1M, b=20M$]{ \includegraphics[width=5.4cm ]{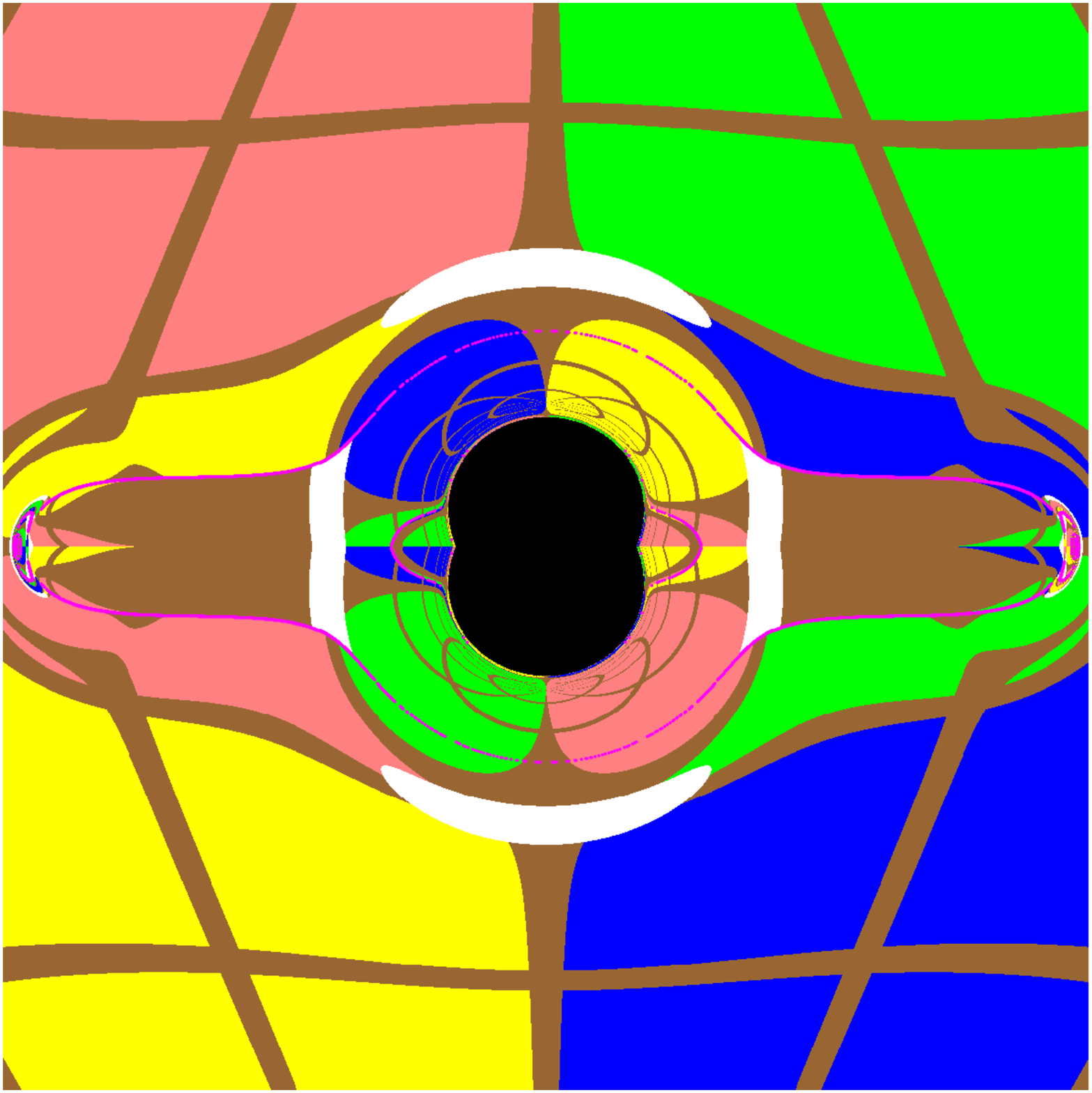}} \subfigure[$\mathcal{M}=1.5M, b=20M$]{ \includegraphics[width=5.4cm ]{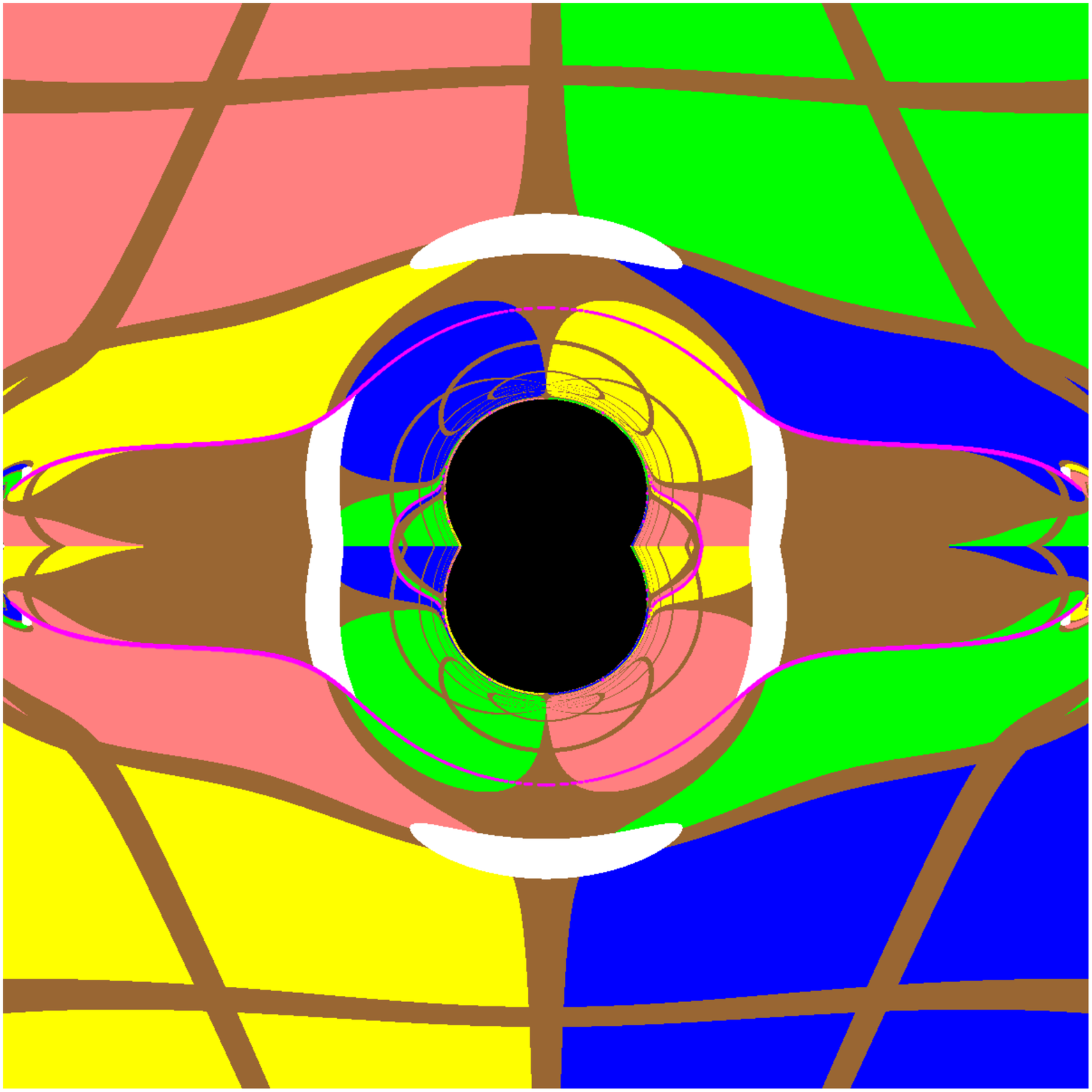}}
\caption{The dependence of black hole shadow on the ring mass $\mathcal{M}$ for fixed Weyl radius $b$. The upper row is for $b=10M$ and The bottle one is for $b=20M$. Here we set the black hole $M=1$ and the observer locates in the position with $r_{obs}=50M$  and  $\theta_{obs}=\pi/2$.}
\label{10}
\end{figure}
With these quantities, one can obtain the coordinates of photon's image in observer's sky \cite{sw,swo,astro,chaotic,binary,sha18,my,sMN,BI,swo7,sb10}
\begin{eqnarray}
\label{xd1}
x&=&-r\frac{p^{\hat{\phi}}}{p^{\hat{r}}}|_{(r_{obs},\theta_{obs})}
=-\sqrt{\frac{r-2M}{r\sin^{2}\theta}}e^{2\nu_{BW}-\lambda_{ext}}\frac{L_{z}}{\dot{r}}|_{(r_{obs},\theta_{obs})}, \nonumber\\
y&=&r\frac{p^{\hat{\theta}}}{p^{\hat{r}}}|_{(r_{obs},\theta_{obs})}=
\sqrt{r^{3}(r-2M)}\frac{\dot{\theta}}{\dot{r}}|_{(r_{obs},\theta_{obs})},
\end{eqnarray}
where the spatial position of observer is set to ($r_{obs}, \theta_{obs}$).
\begin{figure}
\subfigure[$b=2M$]{ \includegraphics[width=5.4cm ]{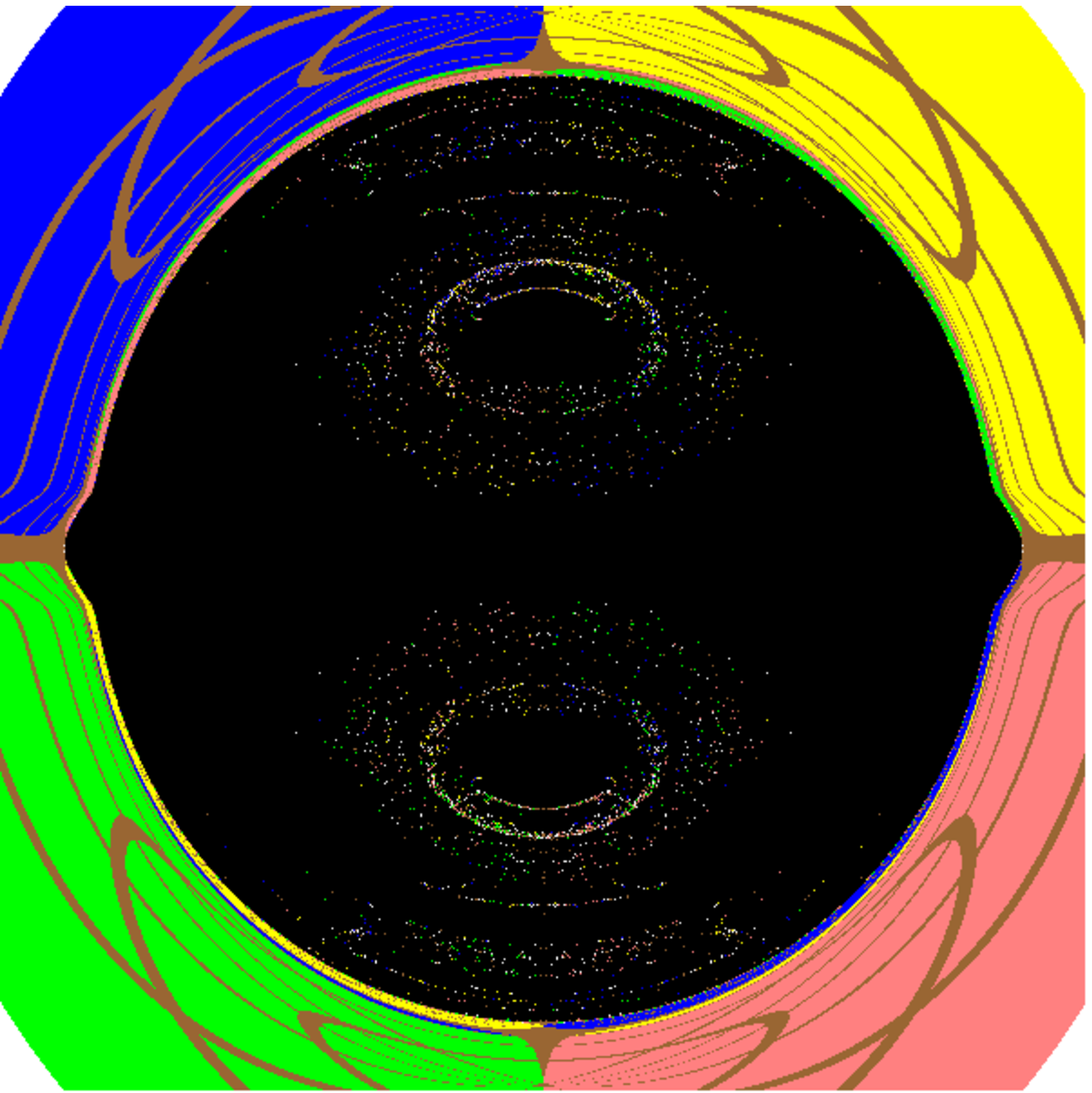}}\subfigure[$b=3M$]{ \includegraphics[width=5.4cm ]{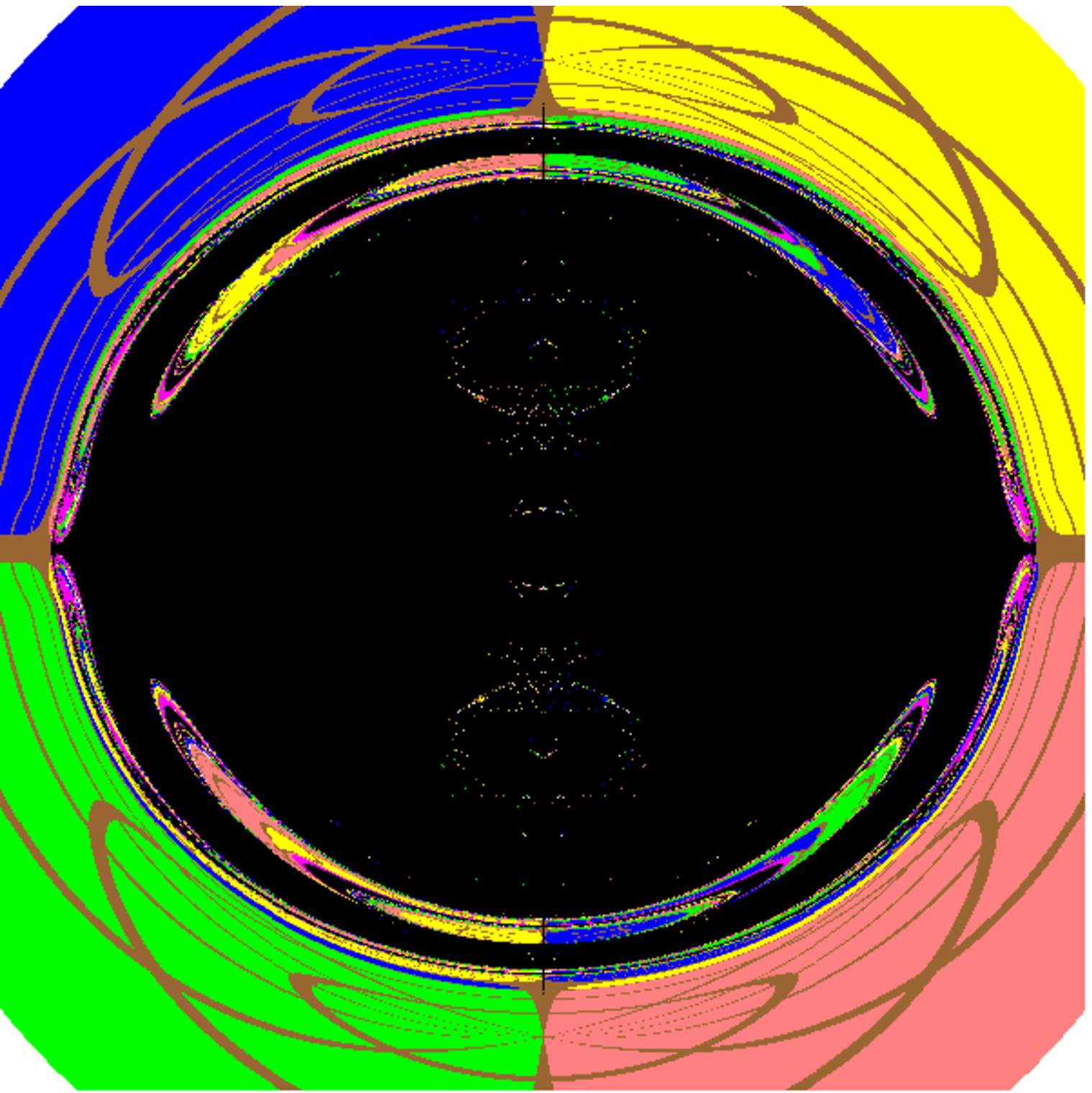}}\subfigure[$b=4M$]{ \includegraphics[width=5.4cm ]{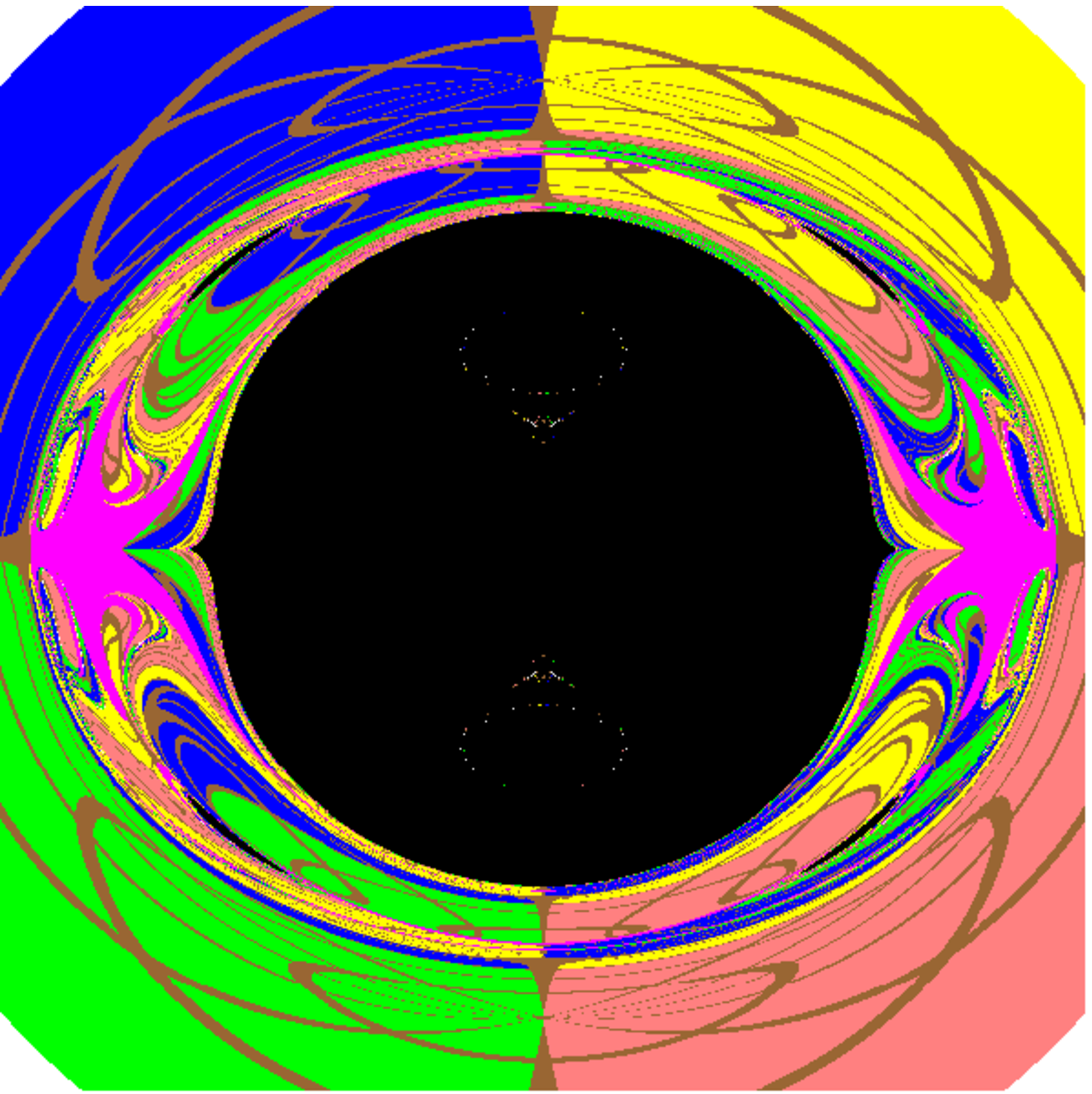}} \subfigure[$b=5M$]{ \includegraphics[width=5.4cm ]{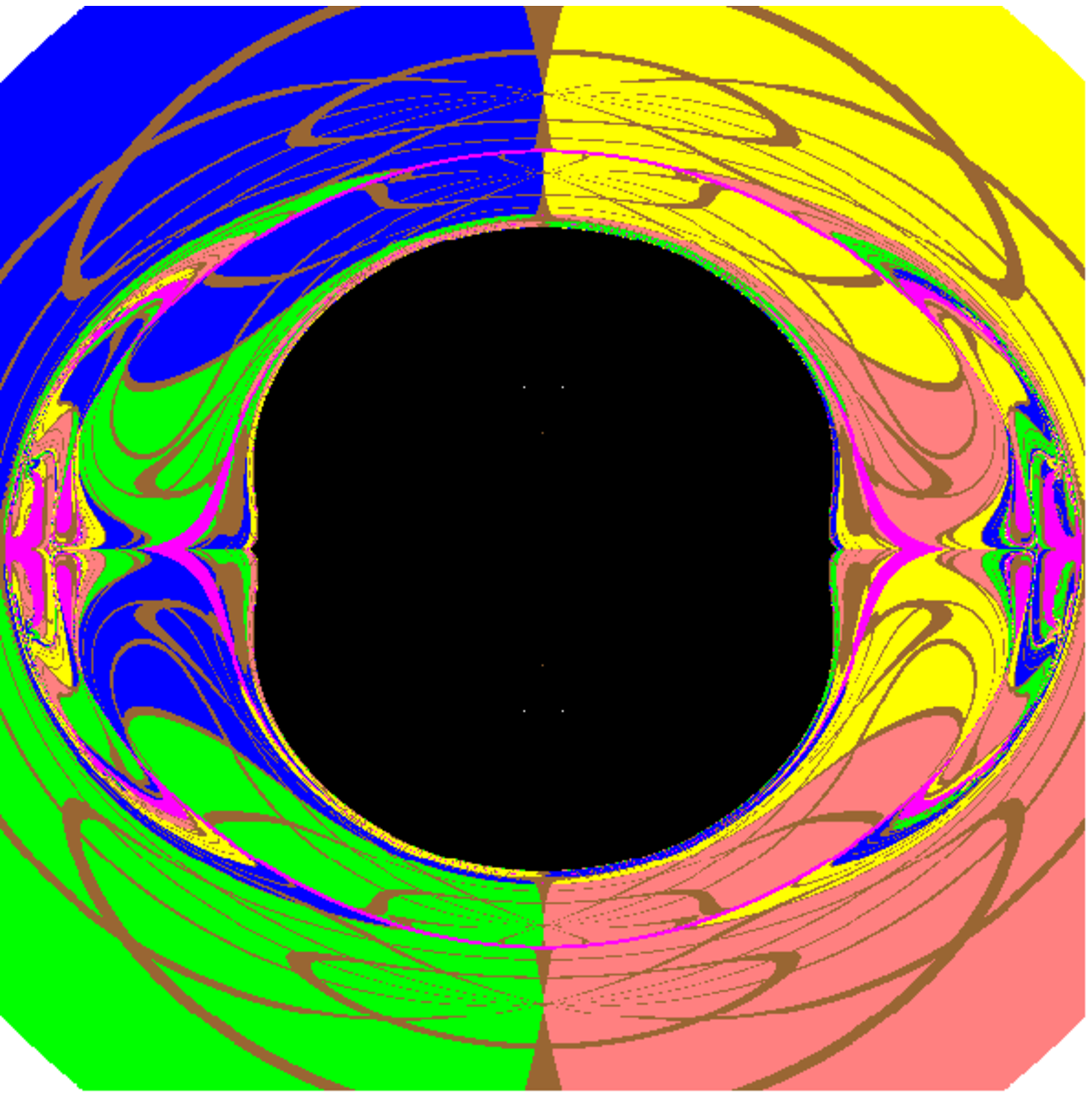}}\subfigure[$b=10M$]{ \includegraphics[width=5.4cm ]{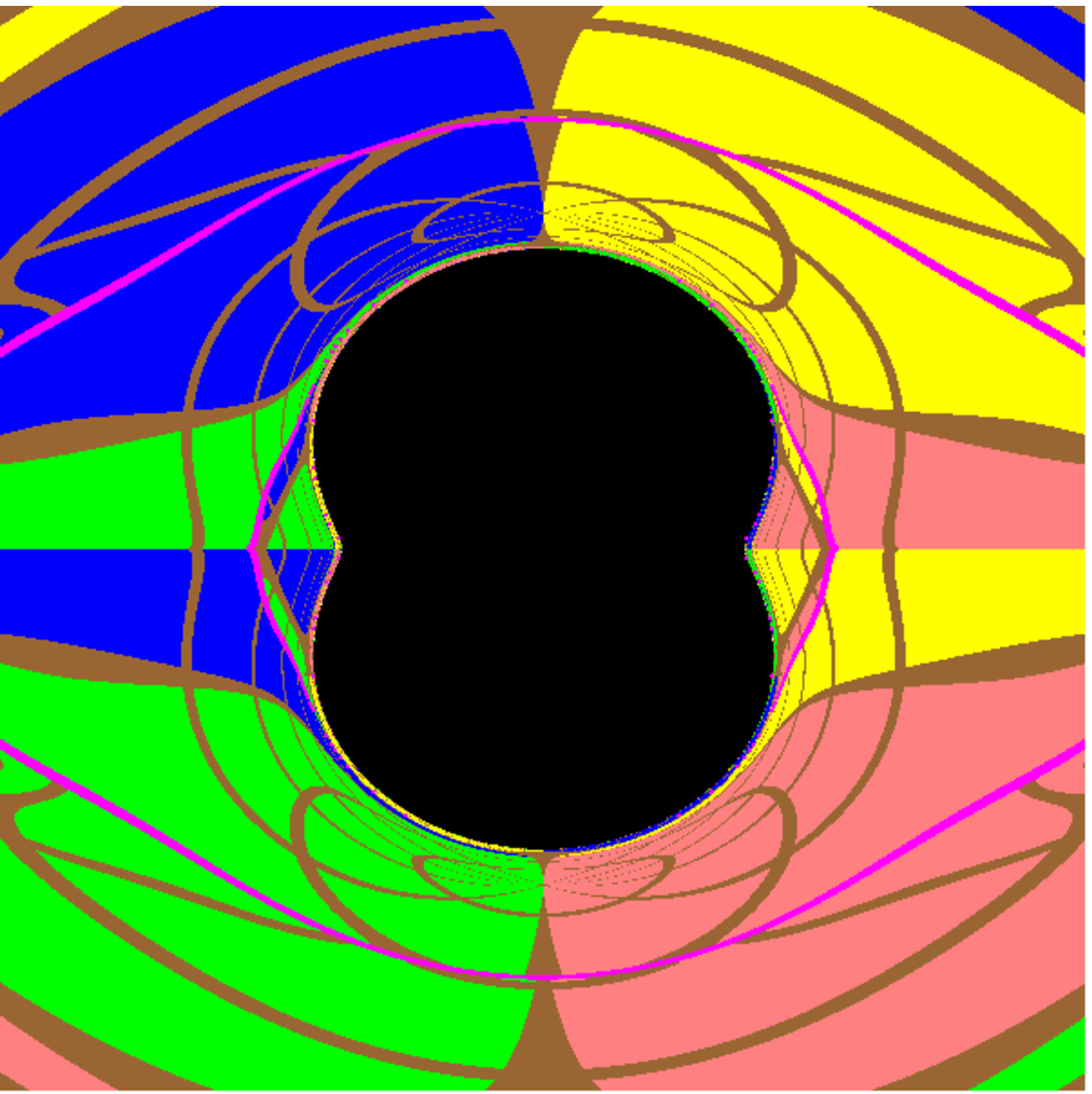}}\subfigure[$b=20M$]{ \includegraphics[width=5.4cm ]{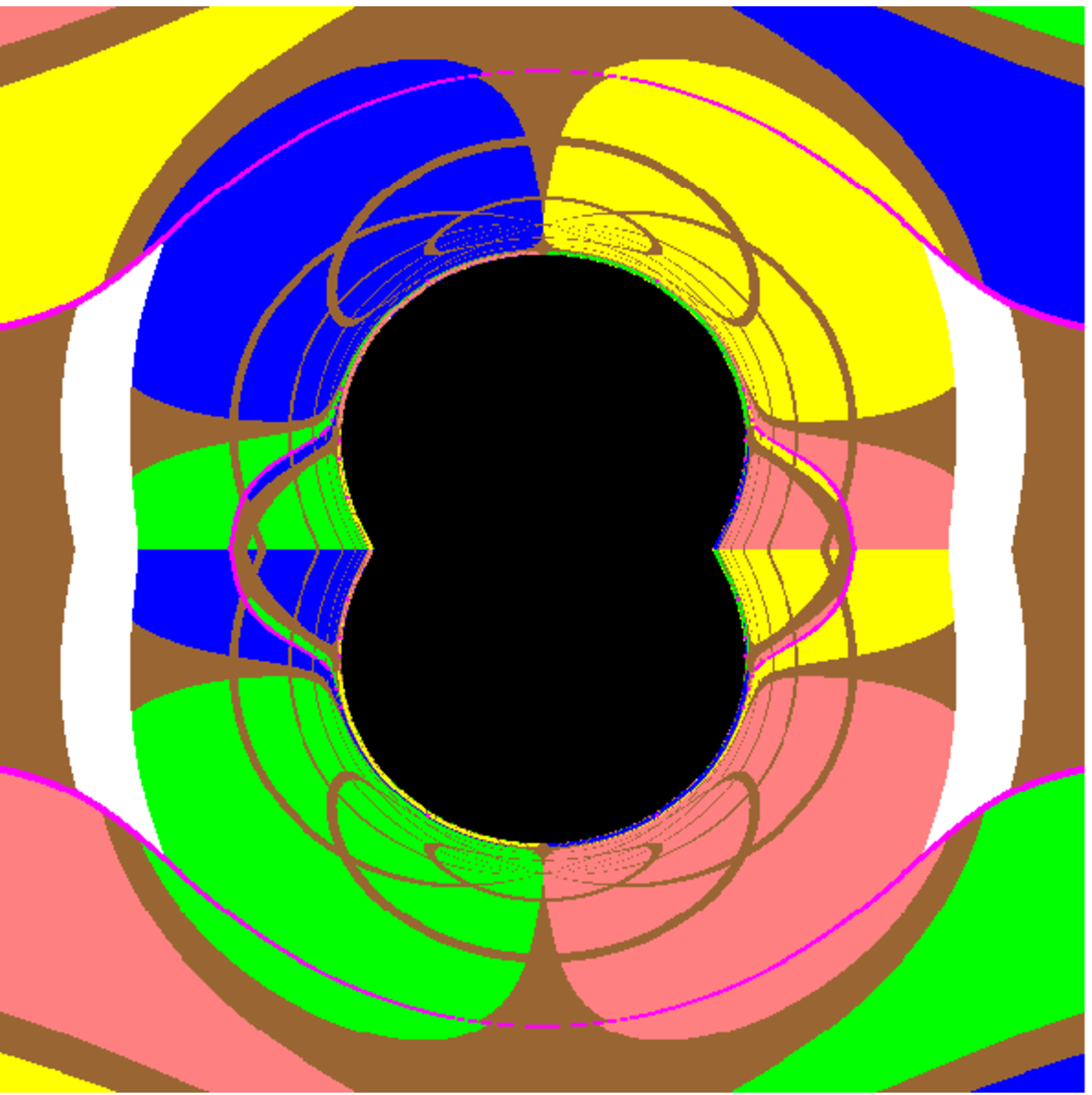}}
\caption{The change of black hole shadow on the Weyl radius $b$. Here we set the ring mass $\mathcal{M}=1.5M$ and the black hole $M=1$, and the observer locates in the position with $r_{obs}=50M$  and  $\theta_{obs}=\pi/2$.}
\label{15b}
\end{figure}

In Fig.\ref{10}, we present the dependence of the black hole shadow on the mass of Bach-Weyl ring for the fixed Weyl radius. Here we assume that Bach-Weyl ring does not emit light by itself and the observer is located on the equatorial plane of the black hole (the inclination angle $\theta_{obs}=\pi/2$). With the increase of the ring mass $\mathcal{M}$, we find that the shadow becomes gradually prolate along the axis of symmetry, but turns concave in the equatorial plane, which leads to that the black hole shadow becomes a ``8" type shape in the case with the larger ring mass $\mathcal{M}$. Moreover, the size of the black hole shadow increases with the ring mass $\mathcal{M}$ for different $b$. In Fig.\ref{10}, we also marked the images of Bach-Weyl ring in magenta. It is shown that the image of the Bach-Weyl ring present a flying saucer shape which distributes symmetrically in the both sides of the equatorial plane. With the increase of the ring mass $\mathcal{M}$, the saucer shape of the image becomes wider for the Bach-Weyl ring.  The presence of Bach-Weyl ring also affects the shape of Einstein ring. Especially, in the cases where the ring mass is set to $\mathcal{M}=M$ or $1.5M$ and the Weyl radius is set to $b=10M$ or $20M$, the Einstein ring is broken into four parts, and then the Einstein ring turns into the so-called Einstein cross. From Fig.\ref{10}, due to the combined action of the black hole and the Bach-Weyl ring, for some selected $b$,  some parts of Einstein cross lie inside Bach-Weyl ring and the other is located outside the ring.

In Fig.\ref{15b}, we show the change of the black hole shadow on the Weyl radius $b$ of Bach-Weyl ring for fixed ring mass $\mathcal{M}$. When the Weyl radius $b$ is small, the shadow is a oblate silhouette and it is convex in the equatorial plane, but there exist some bright dispersion points in the black hole shadow, which possesses self-similar fractal structures originating from the chaotic lensing. With the increase of $b$, chaotic lensing becomes more distinct so that some bright strips appear in the black hole shadow, which yields the emergence of some eyebrow-shape shadows. With the further increase of $b$, the bright dispersion points with self-similar fractal structures disappears and black hole shadow becomes gradually concave in the equatorial plane. Finally, the black hole shadow becomes a prolate silhouette with the ``8" shape  again in the cases with the larger $b$. Moreover, the size of the black hole shadow decreases with the Weyl radius $b$.
\begin{figure}
\subfigure[$\theta_{obs}=0^{\circ}$]{ \includegraphics[width=5.6cm ]{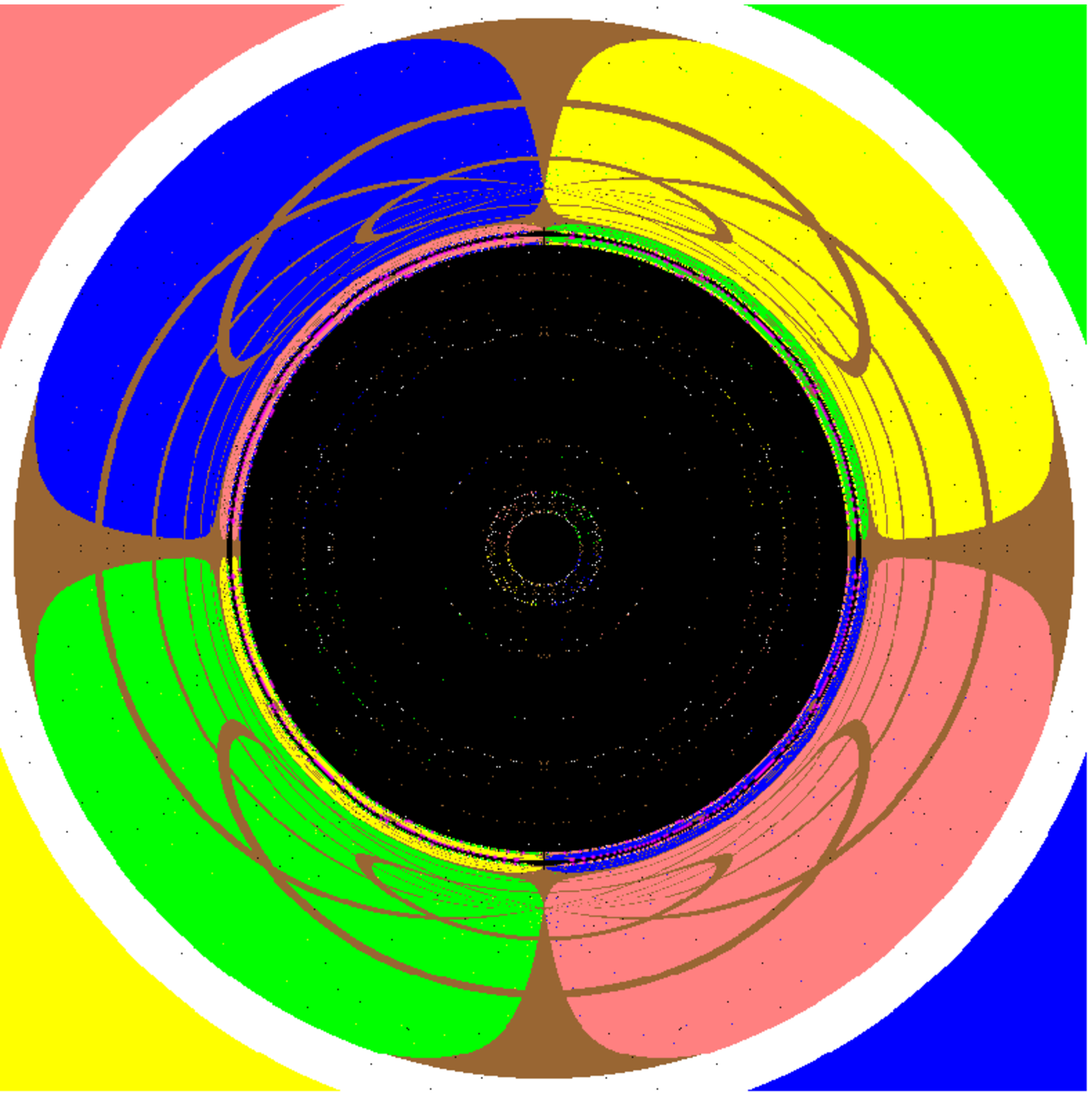}}\subfigure[$\theta_{obs}=30^{\circ}$]{ \includegraphics[width=5.6cm ]{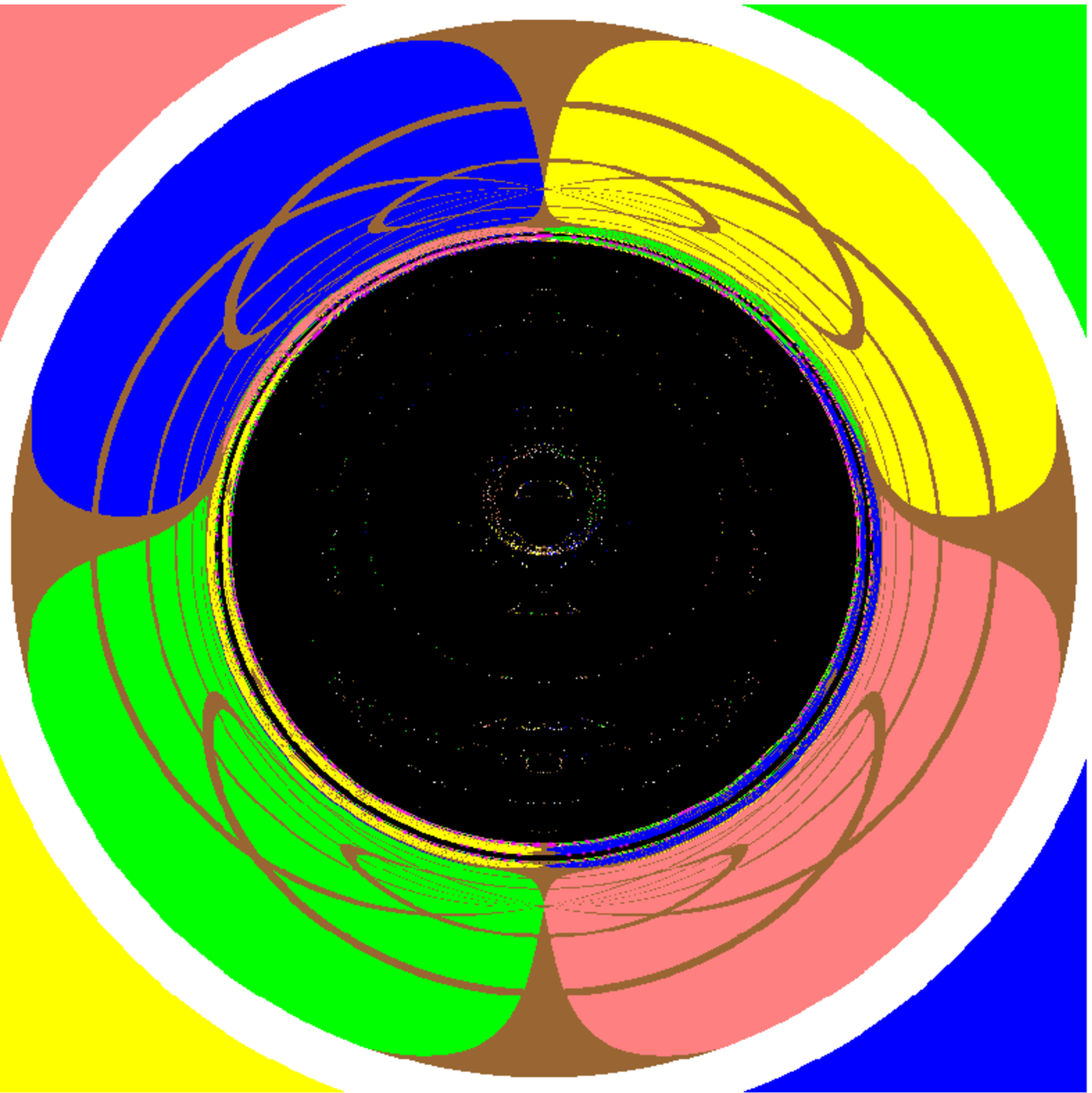}} \subfigure[$\theta_{obs}=60^{\circ}$]{ \includegraphics[width=5.6cm ]{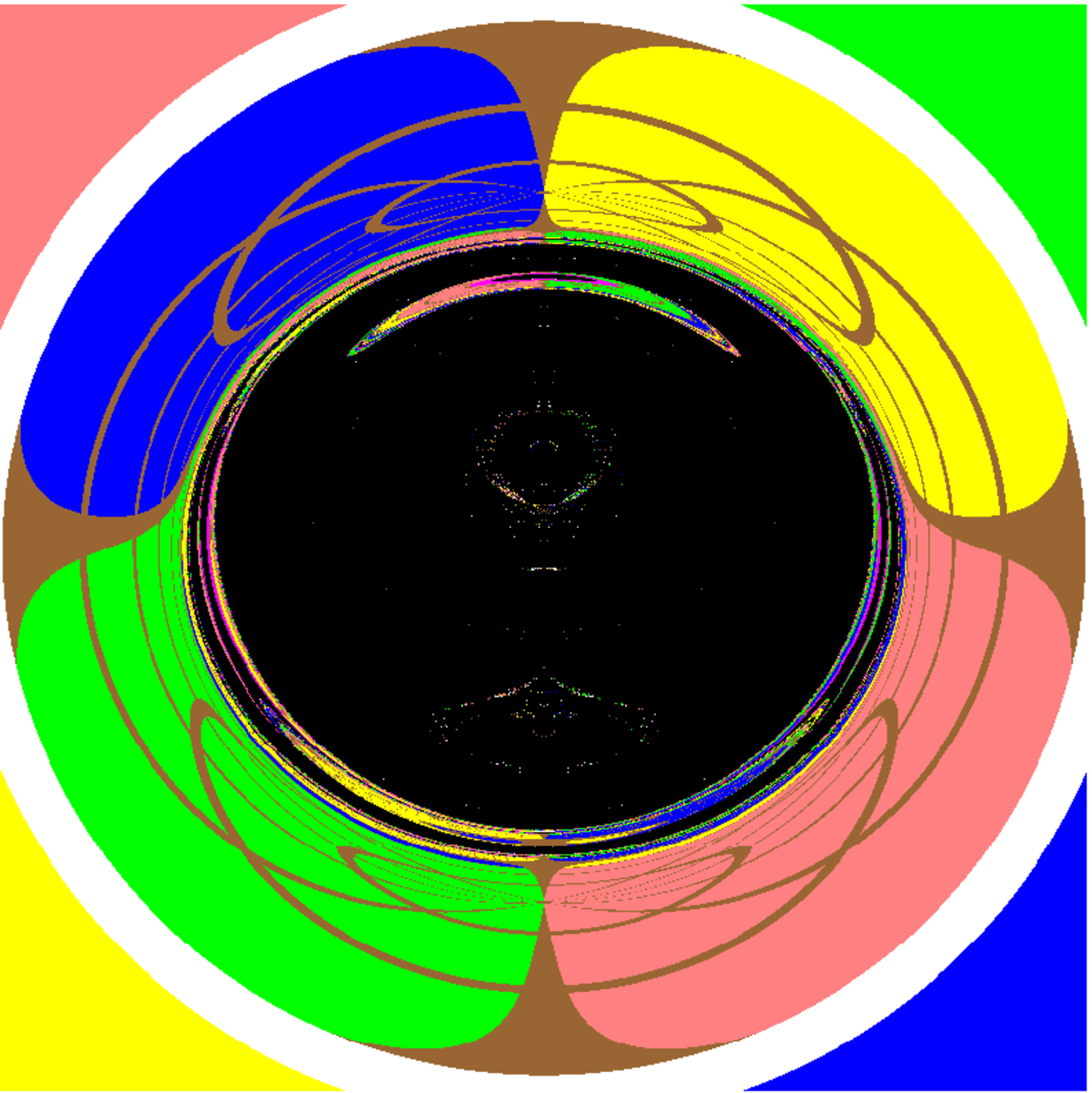}}\subfigure[$\theta_{obs}=90^{\circ}$]{ \includegraphics[width=5.6cm ]{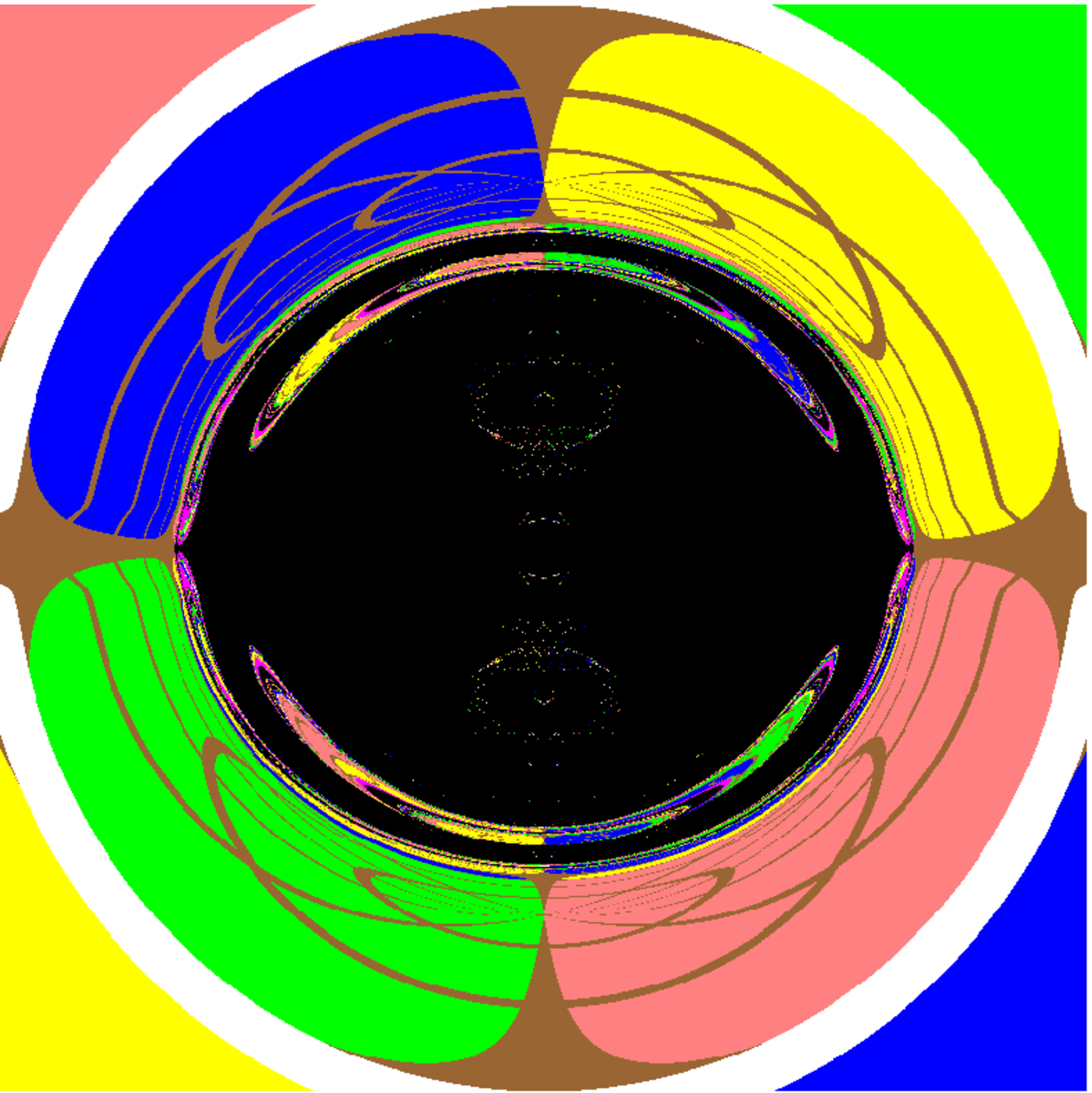}}
\caption{The shadows of Schwarzschild black hole surrounded by Bach-Weyl ring for the observers with different inclination angle. Here we set $\mathcal{M}=1.5M$ and Weyl radius $b=3M$.}
\label{jd}
\end{figure}

In Fig.\ref{jd}, we plot the shadows for the observers with different inclination angle. As $\theta_{obs}=0^{\circ}$, it is obvious that the shadow is center symmetric. Moreover, we find that
there are concentric bright rings imbedded in the black disc, which is qualitatively different from that in the case of usual spherical black hole. These distinct features in the shadow can be attributed to
the effect of Bach-Weyl ring on the spacetime structures.  For the observer with the other inclination angle $\theta_{obs}$, the center symmetry of shadow is broken and the shadows becomes axially symmetric. With the increase of inclination angle $\theta_{obs}$, we find the eyebrow shape shadows appears more distinctly. And finally, as $\theta_{obs}=90^{\circ}$, the pattern of shadow owns the symmetry along the equatorial plane.

\section{Invariant phase-space structures and the shadow cast by a Schwarzschild black hole surrounded by a Bach-Weyl ring}

\begin{figure}[ht]
\center{\includegraphics[width=8cm ]{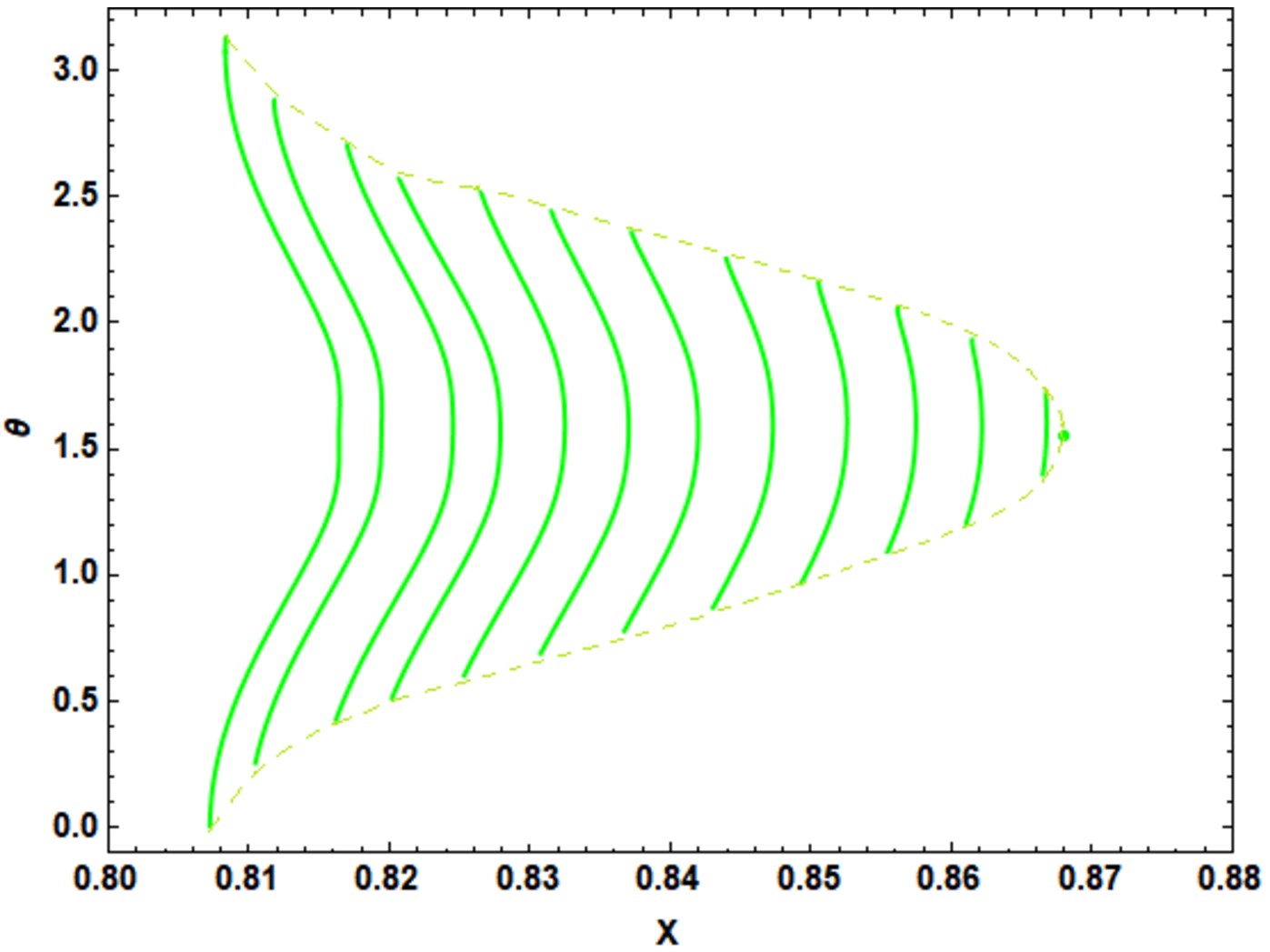}
\caption{Light ring (dot) and the families of periodic Lyapunov orbits (solid line) in the spacetime of a Schwarzschild black hole surrounded by a Bach-Weyl ring with $\mathcal{M}=1.5M$ and Weyl radius $b=3M$.}
\label{bwzqt}}
\end{figure}
\begin{figure}
\subfigure[]{ \includegraphics[width=6.5cm ]{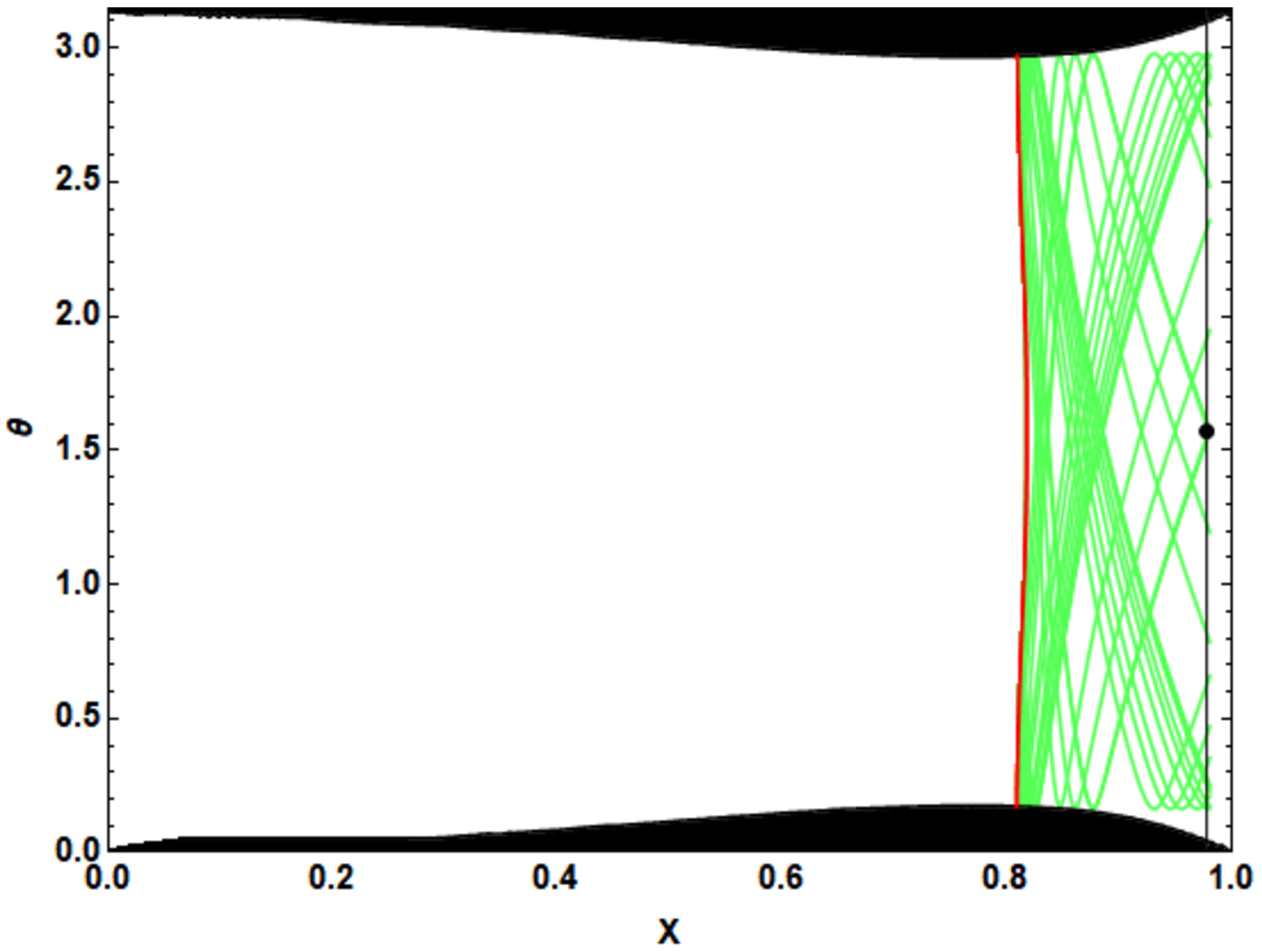}}\subfigure[]{ \includegraphics[width=5cm ]{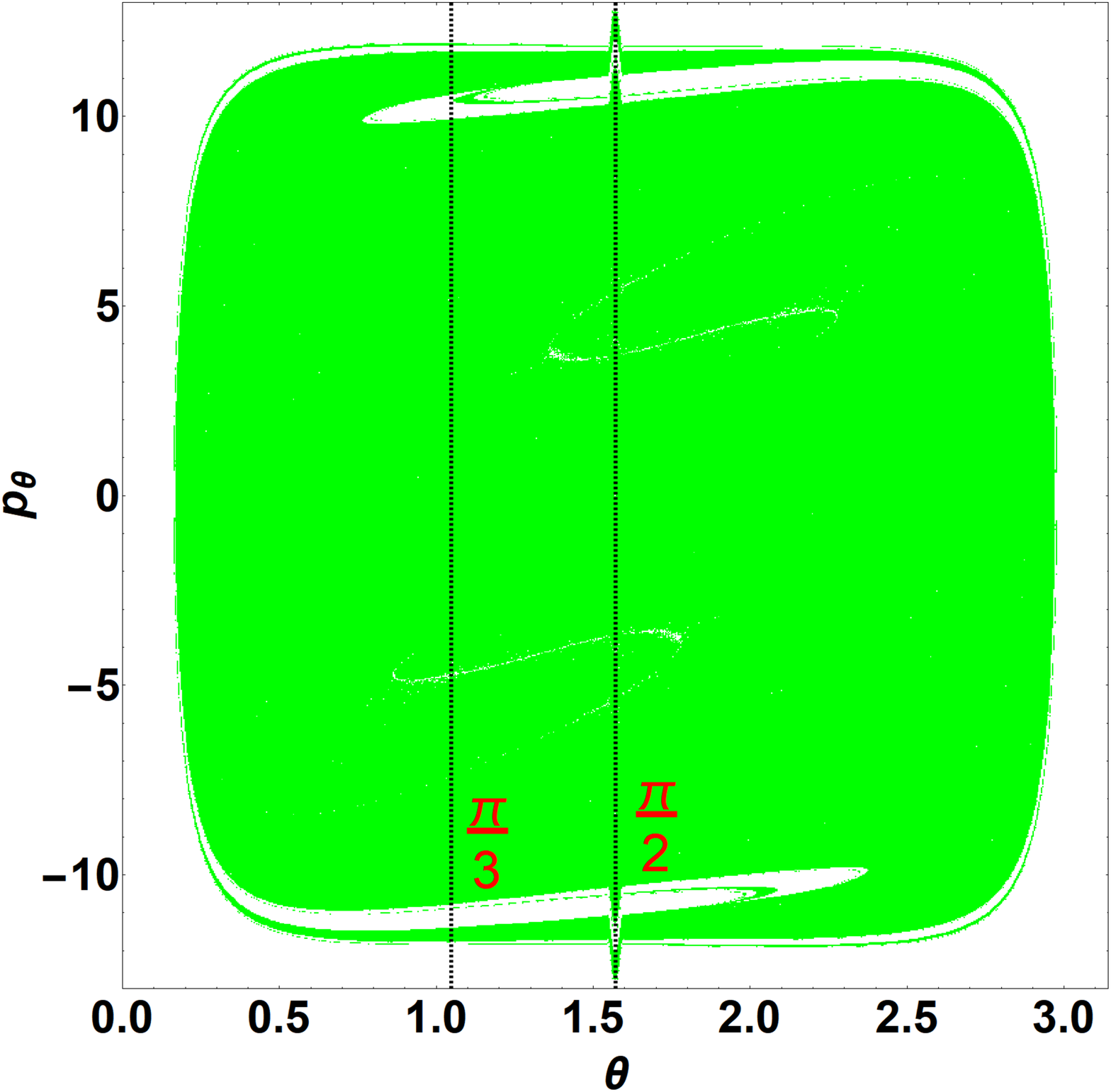}} \subfigure[$\theta_{obs}=60^{\circ}$]{ \includegraphics[width=5.6cm ]{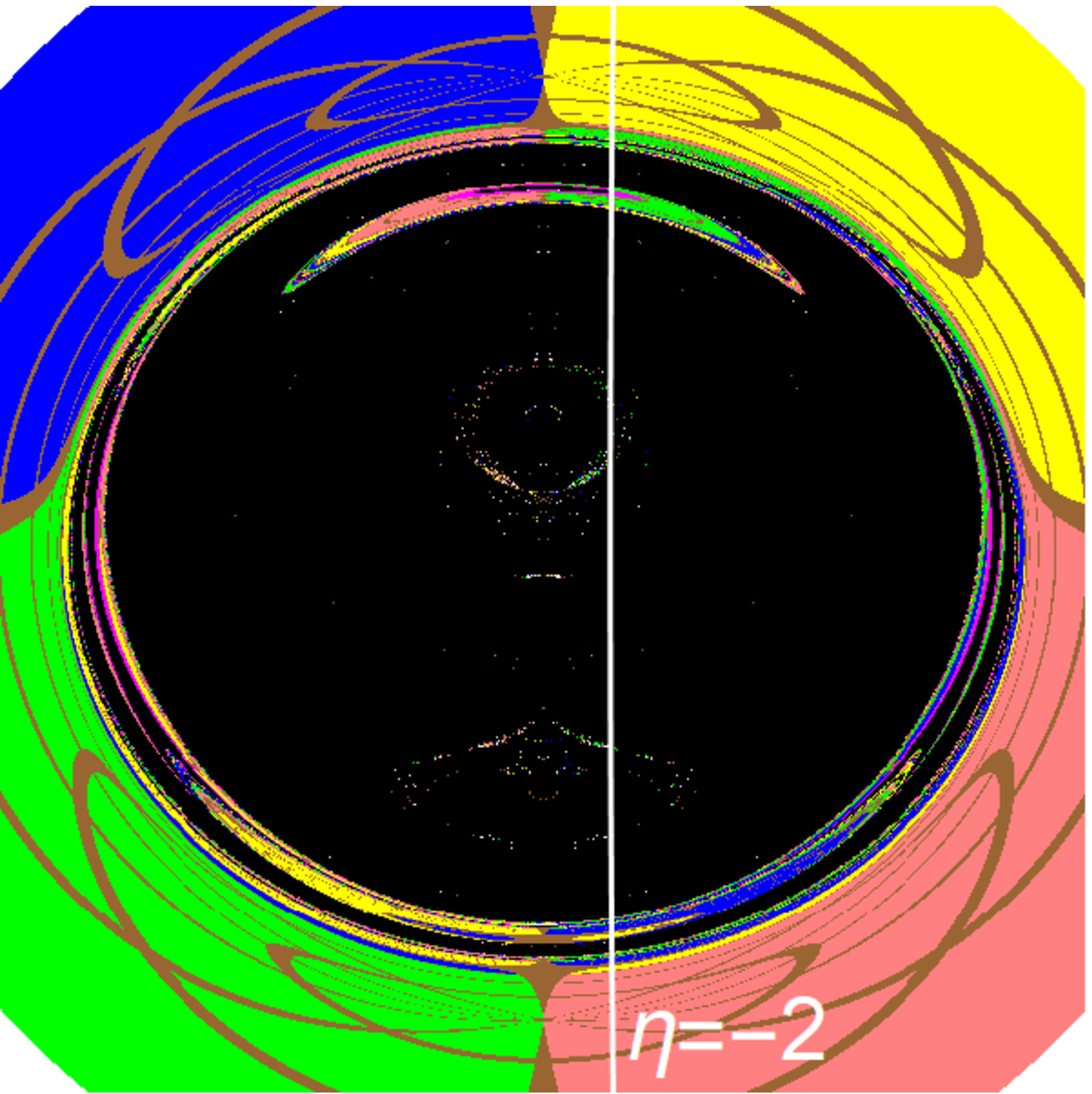}}\subfigure[$\theta_{obs}=90^{\circ}$]{ \includegraphics[width=5.6cm ]{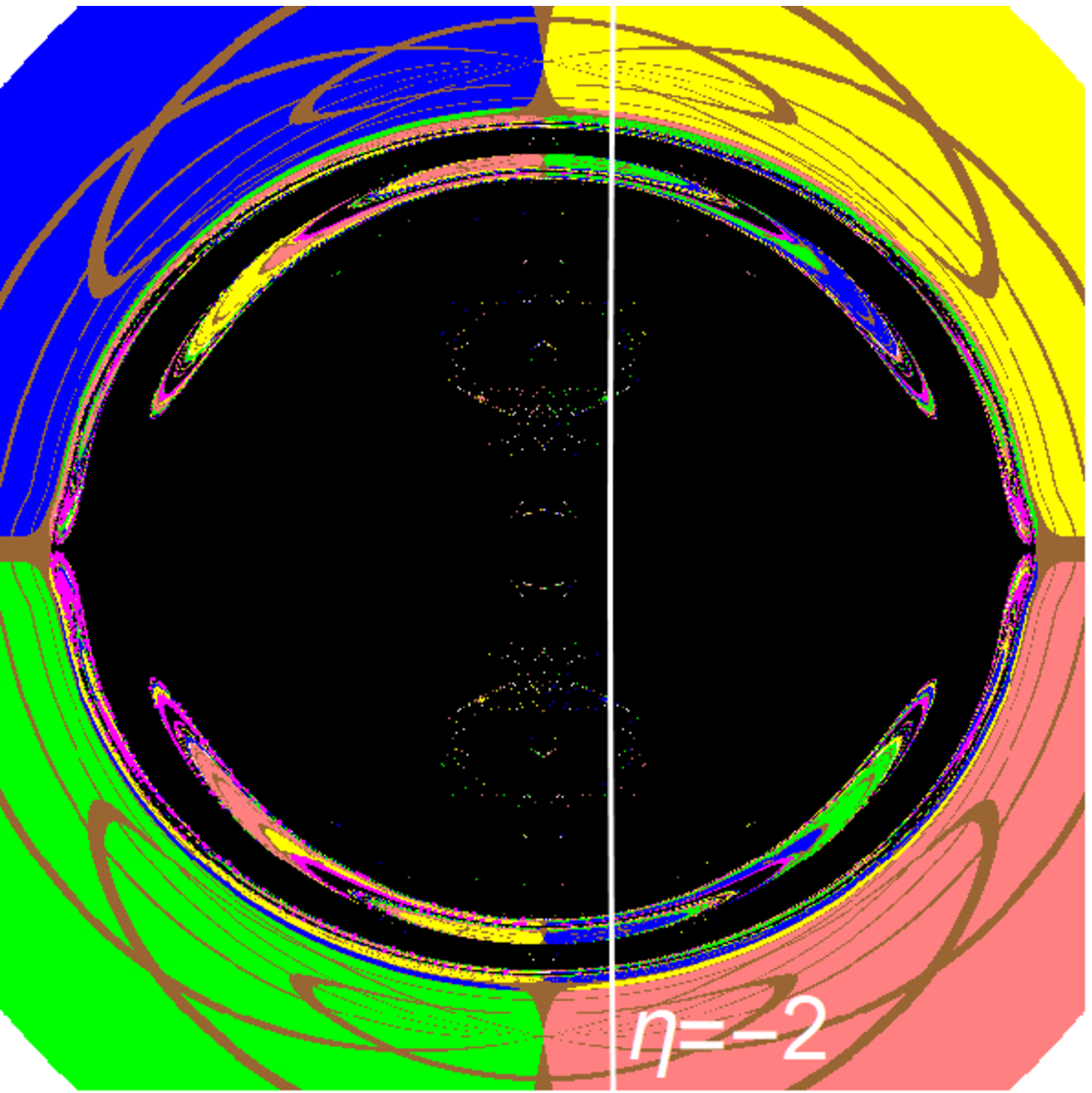}}
\caption{(a) Projection of the unstable invariant manifold (green lines) associated with the periodic orbits (red lines) for $\eta=-2$. The dark regions are the forbidden regions for photons and the black dot represents observer. (b) Poincar\'{e} section in the plane ($\theta, p_{\theta}$) for the unstable manifold (green) with $\eta=-2$ at the observers  position. The angle $\theta$ ($60^{\circ}$ and $90^{\circ}$) marked near two dash lines denote the inclination angle of observers. (c) and (d) correspond to the shadows observed by the observer with the inclination angle $\theta_{obs}=60^{\circ}$ and $\theta_{obs}=90^{\circ}$, respectively. The white line is for fixed $\eta=-2$.}
\label{jt}
\end{figure}

Let us now in position to analyze the invariant phase-space structures of the photon system in the spacetime of a Schwarzschild black hole with a Bach-Weyl ring and then further explain the formation of the black hole shadow.

The invariant phase space structures contain fixed points, periodic orbits and invariant manifolds. As one of the most important features for dynamical systems,  the invariant phase space structures are applied extensively in the design of space trajectory for various of spacecrafts including a low energy transfer from the Earth to the Moon and a ``Petit Grand Tour" of the moons of Jupiter  \cite{BI17,BI18,BI19,BI20,BI22}. Recently, it is shown that
the invariant phase space structures play an important role in the formation of black hole shadows \cite{BI,my}.

For the spacetime of a Schwarzschild black hole surrounded by a Bach-Weyl ring (\ref{bydg}), the fixed point $x_0=(r_{0},\theta_{0},0,0)$ in phase space $(r,\theta,p_r,p_{\theta})$ is defined by the condition
\begin{eqnarray}
\label{bdd}
\dot{x}^{\mu}=\frac{\partial H}{\partial p_{\mu}}=0,\;\;\;\;\;\;\;\;\;\;\;\;\;\;
\dot{p}_{\mu}=-\frac{\partial H}{\partial x^{\mu}}=0,
\end{eqnarray}
which leads to
\begin{eqnarray}
\label{bdd1}
 V\bigg|_{r_{0},\theta_{0}}=0,\;\;\;\;\;\;\;\;\;\;\;\;\;\;\frac{\partial V}{\partial r}\bigg|_{r_{0},\theta_{0}}=0,\;\;\;\;\;\;\;\;\;\;\;\;\;\;
\frac{\partial V}{\partial \theta}\bigg|_{r_{0},\theta_{0}}=0.
\end{eqnarray}
Linearizing the equations (\ref{bdd}),
\begin{eqnarray}
\label{xxh}
\mathbf{\dot{X}}=J\mathbf{X},
\end{eqnarray}
one can analyze the local stability of the fixed point $x_0=$($r_{0},\theta_{0},0,0$). Here the vector $\mathbf{X}\equiv(\tilde{x}^{\mu},\tilde{p}_{\mu})$ and $J$ is the Jacobian.
The fixed points of the dynamics for the  photon motion are light rings, which are the circular photon orbits in the equatorial plane \cite{BI,fpos2}. Considering that the function $\lambda$ in the metric (\ref{bydg}) is not analytical, we here adopt the case with $\mathcal{M}=1.5M$ and $b=3M$ as an
example to analyze numerically the formation of the shadow of a Schwarzschild black hole with a Bach-Weyl ring (\ref{bydg}) which is shown in Fig.\ref{15b} (b). Solving equation (\ref{bdd1}) numerically, we find that there exist two fixed points in this special case. The positions of these two fixed points in phase space are overlapped  at ($6.869,\pi/2,0,0$), but their impact parameters are  $\eta_1=-14.29$ and $\eta_2=14.29$, respectively.  This distribution of two fixed points is determined by a fact that the Schwarzschild black hole with a Bach-Weyl ring (\ref{bydg}) is a non-rotating spacetime. The further analysis indicates that the eigenvalues of the Jacobian in the equation (\ref{xxh}) are  $\lambda_0=\pm 0.42779$, $\nu=\pm 0.42779\; i$. From Lyapunov central theorem, one can find that each purely imaginary eigenvalue yields a Lyapunov family \cite{BI}, which is  a one parameter family $\gamma_{\epsilon}$ of periodic orbits and the orbit $\gamma_{\epsilon}$ collapses into the fixed point as $\epsilon\rightarrow0$.
In Fig. \ref{bwzqt}, we show the Lyapunov family for the fixed points (light rings)  in the plane ($X,\theta$), where $X$ is a compacted radial coordinate defined as $X=\sqrt{r^{2}-r_{h}^{2}}/(1+\sqrt{r^{2}-r_{h}^{2}})$ as in Ref. \cite{BI}.  The thick dot represent the light ring, and the solid lines denote the corresponding  family of periodic Lyapunov orbits arising from these light rings. These periodic orbits can be parameterized by the impact parameter $\eta=L_{z}/E$ on the interval $[-14.29, 14.29]$. The periodic orbits with the same absolute value of $\eta$  is overlapped in the plane ($X,\theta$) because that the spacetime(\ref{bydg}) is static and axially symmetric.
All of these periodic Lyapunov orbits in Fig. \ref{bwzqt} are nearly spherical orbits with radius $r=6.869$, which play an important role in determining the boundary of shadow of a Schwarzschild black hole with a Bach-Weyl ring (\ref{bydg}). The sign of real eigenvalue $\lambda_0$ determinate the stability of the invariant manifold associated with the fixed point. For the positive real eigenvalue, the invariant manifold  is unstable and  points in the manifold  exponentially approach the fixed point in backward time. The invariant manifolds for each Lyapunov orbit are two dimensional surfaces forming tubes in the three dimensional reduced phase space $(r; \theta; p_{\theta})$. In Fig. \ref{jt} (a), we present a projection of the unstable invariant manifolds ( the green lines) related to the periodic orbit ( the red line) for $\eta=-2$ in the plane ($X,\theta$) for the Schwarzschild black hole with a Bach-Weyl ring with $\mathcal{M}=1.5M$ and $b=3M$. The orbits inside the unstable invariant manifold tube can reach the observer from the region near the event horizon of black hole. Moreover,  the periodic orbit touched the boundary of the black region approaches perpendicularly to the boundary $V(r,\theta)=0$, which is similar to that in Ref.\cite{binary}.
In Fig.\ref{jt}(b), we present the Poincar\'{e} section in the plane ($\theta, p_{\theta}$) for the unstable manifold (green) with $\eta=-2$ at the observers position. The self-similar fractal structures in the unstable manifold are responsible for the self-similar fractal behavior in black hole shadow caused by chaotic lensing. The photons starting within the green regions always move only in the unstable manifold tube. In Fig.\ref{jt} (b),  the intersections of the unstable manifold with the dashed line denotes the trajectories which can be detected by the observer with the corresponding inclination angle. These intersection points also determine the positions of the photons on the image plane and the boundary of the black hole shadow. In Fig. \ref{jt} (c) and (d), we present the lensing images marking the intersection points for fixed $\eta=-2$ with the inclination angle $\theta_{obs}=60^{\circ}$ and $90^{\circ}$, respectively. Comparing subfigure (b) with subfigures (c)and (d) in Fig.\ref{jt}, we find that the bright stripes and bright region in Fig.\ref{jt}(c) and (d)  originate from the white regions in the Poincar\'{e} section,  and the fractal-like structure shown in Fig.\ref{jt} (b) accounts for the fractal shadow structure in Fig.\ref{jd}. These correspondence shows that
the shadow of the Schwarzschild black hole with a Bach-Weyl ring is exactly determined by the unstable invariant manifold associated with the fixed points, which is consistent with those in the cases of Kerr black hole with scalar hair \cite{BI} and of the Bonnor black dihole spacetime \cite{my}.

It is well known that photon sphere plays an important role in the formation of black hole shadow. In the Schwarzschild spacetime, the light rays which do not form the photon sphere are either escape to infinity or get captured by the black hole.  In this sense, the photon sphere can be treated as a basin boundary that separates the basins of escape
and capture of the light rays. Thus, the motion of light rays near the photon sphere are crucial to understanding the properties of black hole shadows.
\begin{figure}[ht]
\center{\includegraphics[width=5cm ]{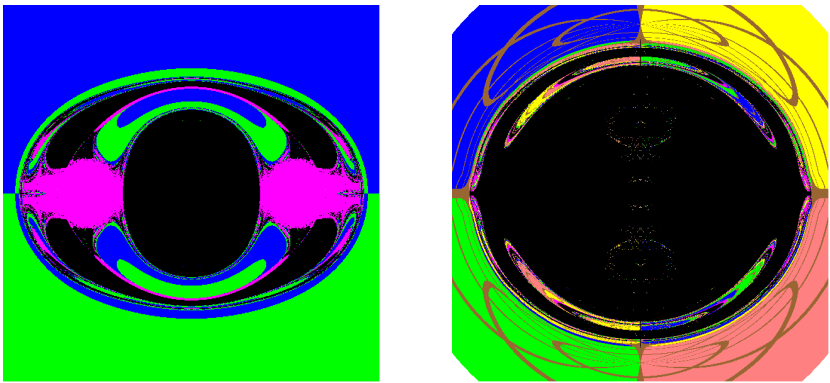}\quad\quad\quad\includegraphics[width=5cm ]{1503.eps}
\caption{Comparison between basins of attraction near the photon sphere ( in the left panel ) and black hole shadow ( in the right panel) at observer for a Schwarzschild black hole surrounded by a Bach-Weyl ring with $\mathcal{M}=1.5M$ and Weyl radius $b=3M$.}
\label{guanqiu0}}
\end{figure}
Recently, Shoom \textit{et al} \cite{shoom1,shoom2} studied the properties of photon sphere of a Schwarzschild black hole distorted by an external gravitational field with a
quadrupole moment. It is shown that there exists a fractal basin boundary due to the chaotic
behavior of null geodesics around the distorted black hole.
In the left panel in Fig.(\ref{guanqiu0}), we present the basins of attraction near the photon sphere for a Schwarzschild black hole surrounded by a Bach-Weyl ring with $\mathcal{M}=1.5M$ and  $b=3M$. Black points correspond to the trajectory got captured by the black hole and magenta points denote the trajectory falling into the Bach-Weyl ring. While, blue and green points correspond to trajectories escaped from the
black hole in the upward and downward directions, respectively. Obviously, the basin boundary near the photon sphere possesses some fractal structures, which is similar to that obtained in the distorted Schwarzschild black hole \cite{shoom1}.
\begin{figure}[ht]
\center{\includegraphics[width=5cm ]{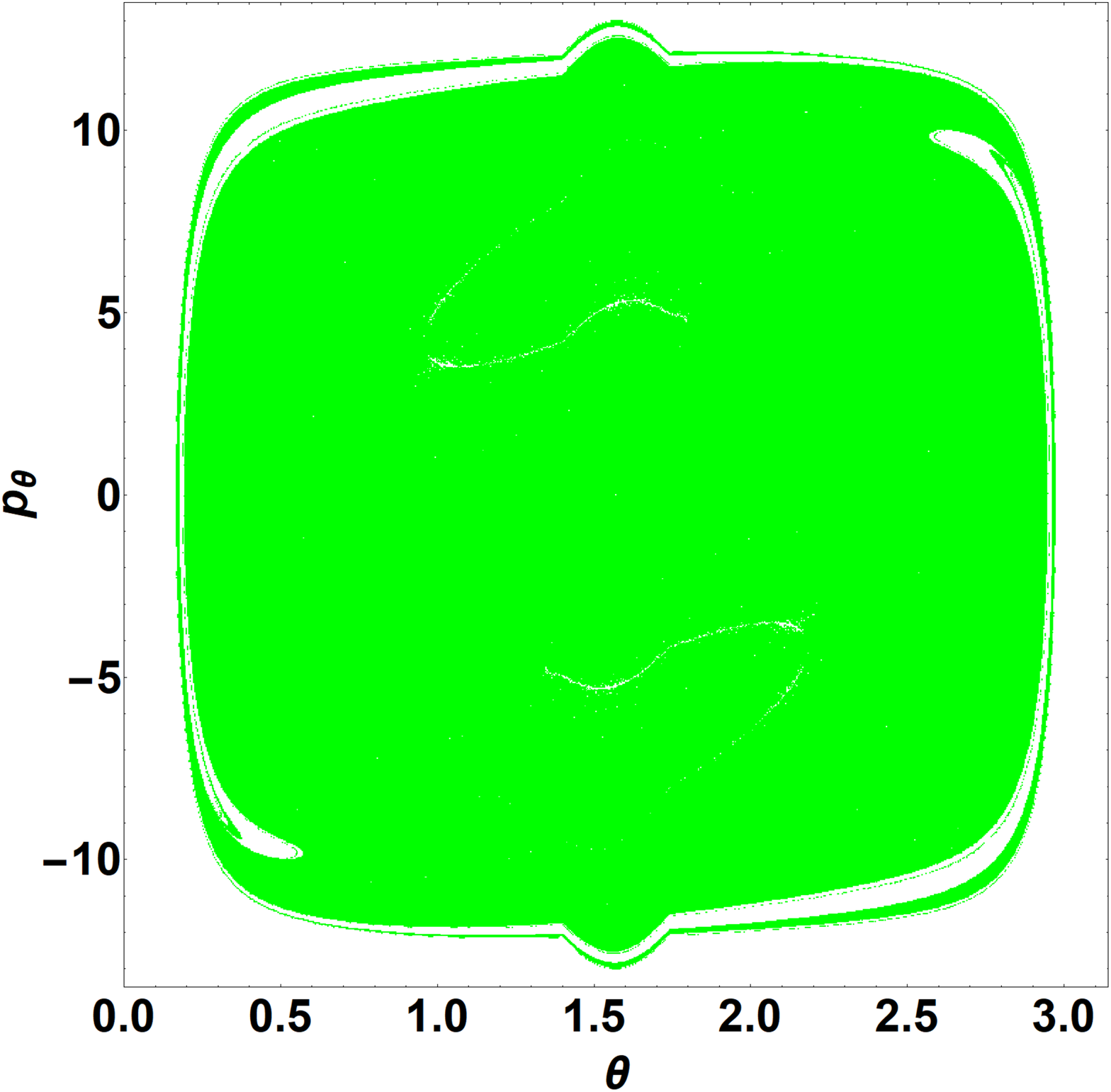}\quad\quad\quad\includegraphics[width=5cm ]{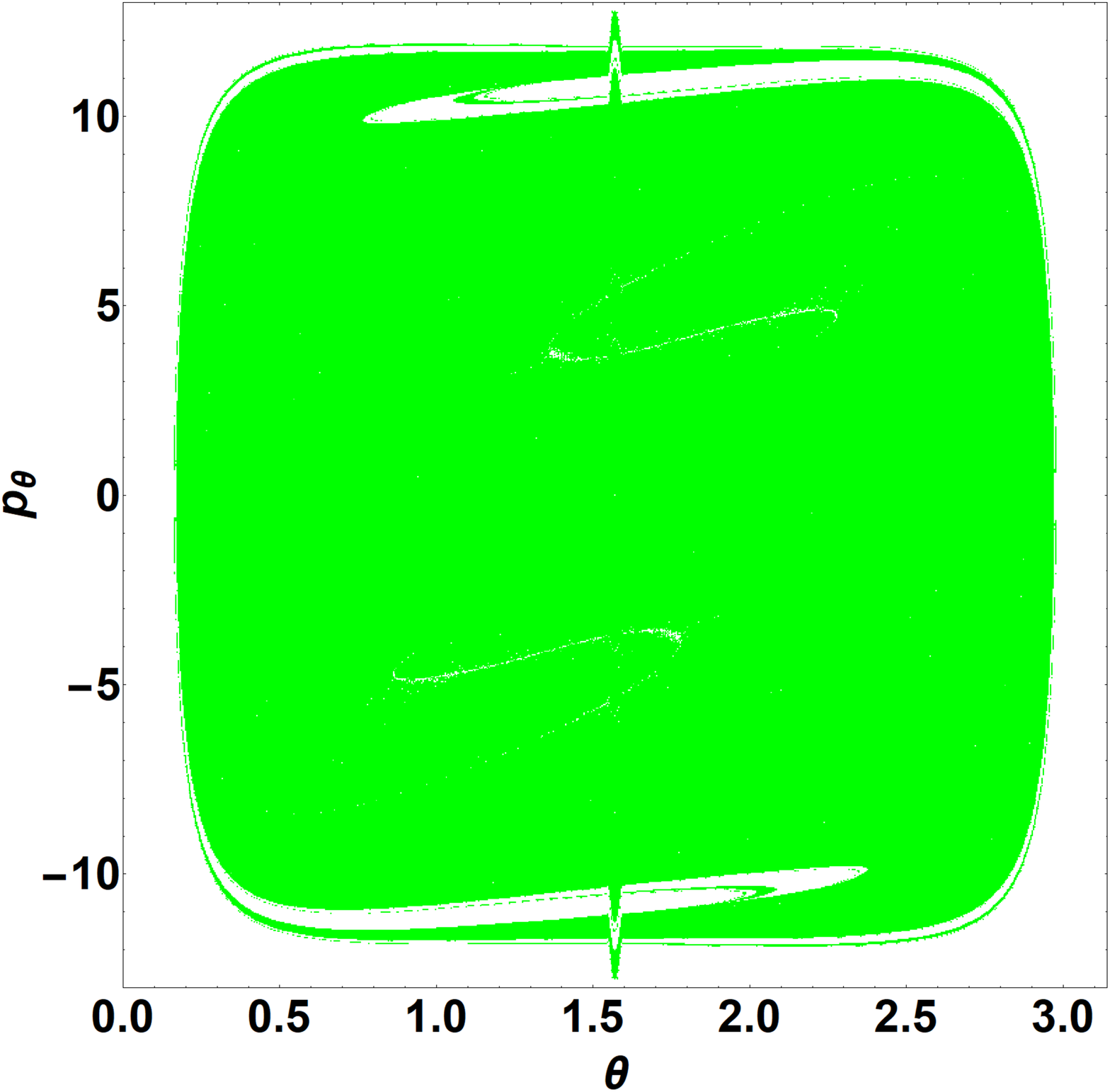}
\caption{ Poincar\'{e} section in the plane ($\theta, p_{\theta}$) for the unstable manifold (green) with $\eta=-2$ in the background of a Schwarzschild black hole surrounded by a Bach-Weyl ring with $\mathcal{M}=1.5M$ and Weyl radius $b=3M$. The left panel is for the section near the photon sphere, and the right panel is for the section at spatial infinite observer. }
\label{gqjm1}}
\end{figure}
From Fig. (\ref{guanqiu0}), we find that basins of attraction near the photon sphere  possess many similar structures to black hole shadow at observer for a Schwarzschild black hole surrounded by a Bach-Weyl ring with $\mathcal{M}=1.5M$ and  $b=3M$. It shows that photon sphere is very important in the formation of black hole shadow. However, we also note that there exist some distinct differences  between the basins of attraction near photon sphere and the black hole shadow. For example, some bright dispersion points appear in the middle of the black hole shadow, but vanish in the basins of attraction near photon sphere. Moreover, the magenta points corresponded to the trajectory falling into the Bach-Weyl ring have the distinct distribution regions in the both cases. The main reason is that
the black hole shadow also depend on the propagation of light ray in the spacetime and the position of observer because the coordinates of photon's image in observer's sky (\ref{xd1}) are determined by the locally measured four-momentum $p^{\hat{\mu}}$ of a photon at the observer. This is also the reason why the shadow radius of a
Schwarzschild black hole  at spatial infinite is $3\sqrt{3}M$ rather than the radius of photon sphere $3M$. From the previous discussion, we know that the black hole shadow is exactly determined by the unstable invariant manifold associated with the fixed points. In Fig. (\ref{gqjm1}), we present Poincar\'{e} sections ($\theta, p_{\theta}$) for the unstable manifold (green)  with $\eta=-2$ near the photon sphere and  at spatial infinite observer, respectively. Although some similar fractal structures appear in both sections, it is obvious that their positions and sizes in the pattern depend on the position of Poincar\'{e} section.
\begin{figure}[ht]
\center{\includegraphics[width=10cm ]{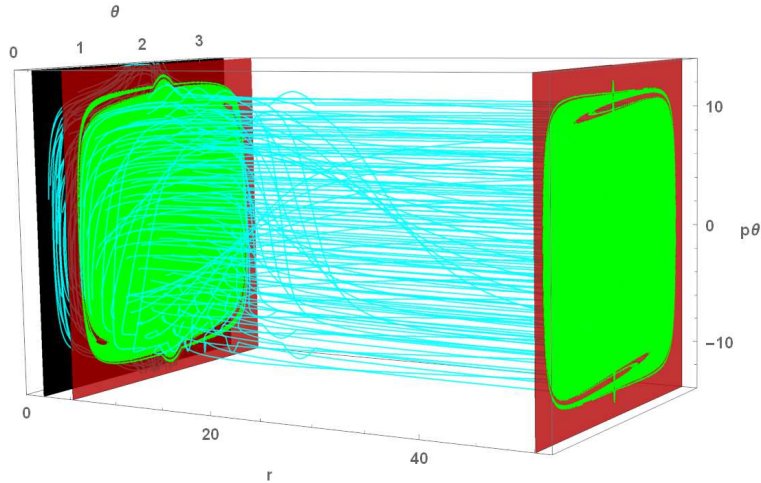}
\caption{Poincar\'{e} sections near the photon sphere  and at spatial infinite observer, and the unstable invariant manifold (light blue curves) between two sections. The black section denotes the event horizon of the black hole. Here, we set $\eta=-2$, $\mathcal{M}=1.5M$ and $b=3M$.}
\label{gqjm2}}
\end{figure}
This is also shown in Fig.(\ref{gqjm2}) in which we present the unstable invariant manifold (light blue line) between Poincar\'{e} sections near the photon sphere and at spatial infinite observer. Generally, the photon orbits in the unstable invariant manifold are irregular curves rather than straight lines, especially in the case the photon motion is chaotic. This leads to the difference between the basins of attraction near the photon sphere and the black hole shadow.
In other words, black hole shadow depends on not only the behavior of light rays near the photon sphere, but also the propagation of photons between the photon sphere and the observer.

\section{Summary}

In this paper we have studied the shadows of a Schwarzschild black hole surrounded by a Bach-Weyl ring. The presence of Bach-Weyl ring leads to that the photon dynamical system is non-integrable and then chaos would appear in the photon motion. Making use of the backward ray-tracing method,
we present numerically the black hole shadow. Our result indicate that
the ring mass and the Weyl radius affects heavily the black hole shadow.
The size of the black hole shadow increases with the ring mass $\mathcal{M}$ and  decreases with the Weyl radius $b$. However, the change of the black hole shadow shape with the ring mass and the Weyl radius becomes more complicated.
With the increase of the ring mass $\mathcal{M}$, we find that the shadow becomes gradually prolate along the axis of symmetry, but turns concave in the equatorial plane, which yields that the black hole shadow becomes a ``8" type shape in the case with the larger ring mass $\mathcal{M}$. In the case with the smaller Weyl radius, the shadow is a oblate silhouette and it is convex in the equatorial plane. Moreover, we also find some bright dispersion points in the black hole shadow, which possesses self-similar fractal structures originating from the chaotic lensing.  With the increase of $b$, the effect chaotic lensing becomes more distinct so that some bright strips appear in the black hole shadow, which yields the emergence of some eyebrow-shape shadows.
But with the further increase of $b$, the bright dispersion points with self-similar fractal structures disappears and black hole shadow becomes gradually concave in the equatorial plane, and finally, the black hole shadow becomes a prolate silhouette with the ``8" shape  again in the cases with the larger $b$. We also study the dependence of the shadows on the inclination angle  of the observers as $\mathcal{M}=1.5M$ and $b=3M$. When $\theta_{obs}=0^{\circ}$,  there are some concentric bright rings imbedded in the black disc, which is qualitatively different from that in the case of usual spherical black hole.  With the increase of inclination angle $\theta_{obs}$, the eyebrow shape shadows appears more distinctly in this case. The presence of Bach-Weyl ring also changes the shape of Einstein ring. For some selected ring mass and Weyl radius, the Einstein ring is broken and then the Einstein cross appears. Due to the combined action of the black hole and the Bach-Weyl ring,  some parts of Einstein cross could lie inside Bach-Weyl ring and the other is located outside the ring.

Moreover, we also study  the change of the image of Bach-Weyl ring with the ring mass and the Weyl radius. In the spacetime of a Schwarzschild black hole with a Bach-Weyl ring, the image of the Bach-Weyl ring present a flying saucer shape, which distributes symmetrically in the both sides of the equatorial plane. With the increase of the ring mass $\mathcal{M}$, the saucer shape of the image becomes wider for the Bach-Weyl ring. Finally, we analyze the invariant manifolds of certain Lyapunov orbits near the fixed point and discuss further the formation of the shadow of a Schwarzschild black hole with Bach-Weyl ring, which implies that the shadow of a Schwarzschild black hole with a Bach-Weyl ring is exactly determined by the unstable invariant manifold associated with the fixed points.

\section{\bf Acknowledgments}

This work was partially supported by the National Natural Science Foundation of China under
Grant No. 11875026, the Scientific Research
Fund of Hunan Provincial Education Department Grant
No. 17A124. J. Jing's work was partially supported by
the National Natural Science Foundation of China under
Grant No. 11875025.

\end{document}